\newlength{\intwidth}

\documentclass{jfm}
\usepackage{amssymb}

\usepackage{array}
\usepackage{amsmath}
\usepackage{latexsym}
\usepackage{bezier}
\usepackage{psfrag,graphicx}
\usepackage{bm}
\usepackage{float}
\usepackage{multirow}
\usepackage{mathrsfs}
\usepackage{color}

\begin{document}

\title[On high Taylor number Taylor vortices]{On high Taylor number Taylor vortices}
\author{Kengo Deguchi}
\affiliation{
School of Mathematics, Monash University, VIC 3800, Australia
}

\maketitle

\begin{abstract}
Axisymmetric steady solutions of Taylor-Couette flow at high Taylor numbers are studied numerically and theoretically. As the axial period of the solution shortens from about \textcolor{black}{one gap length}, the Nusselt number goes through two peaks before returning to laminar flow. In this process, the asymptotic nature of the solution changes in four stages, as revealed by the asymptotic analysis. When the aspect ratio of the roll cell is about unity, the solution \textcolor{black}{captures quantitatively} the characteristics of the classical turbulence regime. Theoretically, the Nusselt number of the solution is proportional to the quarter power of the Taylor number. The maximised Nusselt number obtained by shortening the axial period can reach the experimental value around the onset of the ultimate turbulence regime, although at higher Taylor numbers the theoretical predictions eventually underestimate the experimental values. An important consequence of the asymptotic analyses is that the mean angular momentum should become uniform in the core region unless the axial \textcolor{black}{wavelength} is too short. The theoretical scaling laws deduced for the steady solutions can be carried over to Rayleigh-B\'enard convection.
\end{abstract}

\section{Introduction}

Taylor-Couette flow (TCF) is perhaps among the most studied flows in fluid mechanics. In the 100 years since Taylor's monumental work (Taylor 1923), it has provided an excellent testing ground for theoretical, experimental and numerical studies of rotating shear flows. 
How shear and Coriolis forces alter flow characteristics is important in various applications, \textcolor{black}{and TCF was designed so that} they can easily be adjusted by changing the rotation speed of the inner and outer cylinders. 
%\textcolor{black}{The cylinders' speed is expressed by a dimensionless quantity called the Taylor number.}
Researchers have long been fascinated by the numerous metastable flow patterns observed in the relatively low Taylor number regime (e.g. Andereck et al. 1986). On the other hand, it was only a decade ago that the study of high Taylor number flows became active. Great efforts were made to investigate the nature of turbulence in the parameter space by means of high Taylor number experiments (e.g. Paoletti \& Lathrop 2011; Dennis et al. 2011; \textcolor{black}{van Gils et al. 2011, 2012; Huisman et al. 2012}) and direct numerical simulations (e.g. Ostilla et al. 2013; Brauckmann \& Eckhardt 2013, 2017; Ostilla-M\'onico et al. 2014ab). As summarised in the review paper by Grossmann et al. (2016)\textcolor{black}{, and in fact seen in the pioneering experiments by Lathrop et al. (1992), Lewis \& Swinney (1999),} fully developed turbulence has a surprisingly clean asymptotic character, while there seems to be no definitive Navier-Stokes-based theory to explain it.

%Taylor-Couette flow (TCF) is perhaps among the most studied flows in fluid mechanics. How shear and Coriolis forces alter flow characteristics is important in various applications, and in TCF they can easily be adjusted by changing the rotation speed of the inner and outer cylinders. In the 100 years since Taylor's monumental work (Taylor 1923), TCF has provided an excellent testing ground for theoretical, experimental and numerical studies of rotating shear flows. Numerous patterns have been found in the parameter space and have attracted the interest of many researchers (e.g. Andereck et al. 1986). However, most early studies of TCF focused on relatively low Taylor numbers, and it was only a decade ago that the study of high Taylor number flows became active. Great efforts were made to investigate the nature of turbulence in the parameter space by means of high Taylor number experiments (e.g. Paoletti \& Lathrop 2011; Dennis et al. 2011; van Gils et al. 2012) and direct numerical simulations (e.g. Ostilla et al. 2013; Brauckmann \& Eckhardt 2013, 2017; Ostilla-M\'onico et al. 2014ab). As summarised in the review paper by Grossmann et al. (2016), fully developed turbulence has a surprisingly clean asymptotic character, while there seems to be no definitive Navier-Stokes-based theory to explain it.

This study aims to reveal the asymptotic properties of steady axisymmetric solutions at high Taylor numbers and to compare them with the experimental and numerical results. The analysis of such solutions, known as Taylor vortex solutions, goes back to the weakly nonlinear analysis, for example, by Davey et al. (1962). With modern computational power, it is possible to calculate solutions up to the Taylor number used in the experiments. Of course, the use of Newton's method is essential as the solution is unstable in the high Taylor number regime.

The idea that an unstable solution with a relatively simple structure with respect to time can capture some characteristics of turbulence is not as \textcolor{black}{absurd} as one might think. It is well-known in dynamical systems theory that chaotic dynamics can be approximated by a sufficiently large number of periodic orbits (see Cvitanovi\'c et al. 2012, for example). For moderate Reynolds number shear flows, it was repeatedly reported that a good approximation of the turbulent dynamics can be obtained with a small number of periodic orbits (Kawahara et al. 2012; Willis et al. 2013; Krygier et al. 2021). In phase space, these periodic orbits usually appear with stationary or travelling wave solutions in their vicinity. The advantage of focusing on simple solutions is that their asymptotic nature may be justified theoretically by mathematical analyses. Over the past decade, it has been established that the matched asymptotic expansion is a powerful tool in understanding the behaviour of steady or travelling wave solutions in shear flows (Hall \& Sherwin 2010, Deguchi et al. 2013, Deguchi \& Hall 2014ab, Deguchi 2015, Dempsey et al. 2017, Deguchi \& Walton 2018). Therefore if we are allowed to assume that there is a simple solution that roughly captures the scaling properties of turbulence within the vast phase space, then there is hope for a logical explanation of the scaling from first principles.

In the high Taylor number numerical and experimental studies, the parameter dependence of angular momentum transport was a major focus. The driving force behind those studies was the `analogy' between \textcolor{black}{turbulent} Rayleigh-B\'enard convection (RBC) and TCF. \textcolor{black}{
This analogy was well known at least in the 1960s and has risen and fallen throughout the history of turbulence research (Bradshaw 1969, Dubrulle \& Hersant 2002).
The recent studies have been influenced by Eckhardt et al. (2007) who argued that the phenomenology of RBC turbulence proposed by Grossmann \& Lohse (2000)
can be applied to TCF as well. 
%used the similarity between the angular momentum transport of TCF and the heat flux transport of RBC.
%is based on the fact that the angular momentum transport of TCF can be described by an averaged equation similar to that for the heat flux transport of RBC 
%Combined phenomenology of turbulence and transport equations.
Shortly later the aforementioned high Taylor number TCF experiments confirmed that
 the scaling of the Nusselt number $Nu$ is similar to that observed for RBC by He et al. (2012) ($Nu$ for TCF is defined as the torque on the cylinder wall normalised by its laminar value).
%the ultimate scaling of the Nusselt number $Nu$ observed by He et al. (2012) in very high $Ra$ RBC experiments appears to apply to TCF as well if $Nu$ for TCF
% experiments appears to apply to TCF
%He et al. (2012) in very high $Ra$ RBC.
%
%Somewhat surprisingly, the ultimate scaling of the Nusselt number $Nu$ observed by He et al. (2012) in very high $Ra$ RBC experiments appears to apply to TCF as well if $Nu$ for TCF is defined as the torque on the cylinder wall normalised by its laminar value. 
%when the Rayleigh number $Ra$ is replaced by the Taylor number $Ta$.
} 
It should be remarked that despite the analogy that has been believed, this similarity in the ultimate scaling is actually not at all obvious. As pointed out, for example, by Chandrasekhar (1961), Robinson (1967), Veronis (1970), and Lezius \& Johnston (1976), for the two flows to be equivalent they must be at least axisymmetric. Moreover, the exact equivalence of the two flows \textcolor{black}{requires an infinitesimally narrow cylinder gap, co-rotating cylinders, and a Prandtl number of unity.}

%However, as long as axisymmetry of the flow is assumed in TCF, as in this paper, the structure of the governing equations is very similar to the two-dimensional RBC roll cell problem, and therefore identical scaling may be derived. 
%Although the symmetry-imposed flow may not always resemble the turbulence in the experiment, the analysis may help to understand the analogy.

\textcolor{black}{
It is an interesting question, then, to forget the latter three conditions and ask whether the analogy in the sense of the $Nu$ scaling holds for an axisymmetric Taylor vortex and a two-dimensional roll cell. The equations governing both flows are not the same, but they certainly have a similar structure.
Many researchers have theoretically studied the large Rayleigh number nature of roll cells in RBC over the years;}
%The study of the large Rayleigh number roll cells in RBC has a long history; 
see Pillow (1952), Robinson (1967), Wesseling (1969), Chini \& Cox (2009), Hepworth (2014) for constant Prandtl number flows, Roberts (1979), Jimenez \& Zufiria (1987), and Vynnycky \& Masuda (2013) for asymptotically large Prandtl number flows. Waleffe et al. (2015) and Sondak et al. (2015) shortened the wavelength of the RBC roll cell and found that there is a special wavelength at which the Nusselt number reaches a maximum value. Interestingly, this optimised Nusselt number \textcolor{black}{is} close to that obtained experimentally, although the Rayleigh number they used is much lower than the one used by He et al. (2012). More recently, Wen et al. (2022) continued the same solution branch to higher Rayleigh numbers and claimed that the maximum Nusselt number corresponds to the so-called classical scaling, where the Nusselt number is proportional to the Rayleigh number to the power of one-third (Malkus et al. 1954; Priestley 1954; Grossmann \& Lohse 2000; Kawano et al. 2021). The study by Kooloth et al. (2021) confirmed that roll cells with different wavelengths play important roles in two-dimensionally restricted RBC turbulence. 
This paper is motivated by all the above RBC studies. 

\textcolor{black}{
In the classical turbulence regime of TCF, where Taylor vortices are robustly observed in experiments, the symmetry restriction of the flow may not be necessary for a good agreement between the steady solution and turbulence, and at least a better agreement than in RBC can be expected. The situation is different in the ultimate turbulence regime, where eddies of different sizes and wavelengths are present, making the structure far more complex than classical turbulence. However, as long as the cylinders are not strongly counter-rotating, the experimental observations indicate that vortices of about the scale of the gap are still present, suggesting that Taylor vortices may play some role in the dynamics.
%In the classical turbulence regime of} TCF, the agreement between the steady solution and turbulence would be good even without the symmetry restriction of the flow because Taylor vortices \textcolor{black}{are} robustly seen in experiments.
%expected to remain
%\textcolor{black}{In the ultimate turbulence regime, Taylor vortices often disappear when the time average is taken, but the snapshot shows that vortices of about the scale of the gap are still present unless the cylinders are strongly counter-rotating. 
%Furthermore, ultimate turbulence involves eddies of different sizes and wavelengths, making its structure far more complex than the classical turbulence.
%Not only that, ultimate turbulence involves eddies of different wavelengths and its structure is much more complex than classical turbulence.
}

It should also be noted that previous theoretical studies have shown that analytical approximations can be obtained for vortices with extremely short \textcolor{black}{wavelengths} driven by thermal or Coriolis forces (Hall \& Lakin 1988; Bassom \& Hall 1989; Bassom \& Blennerhassett 1992; Denier 1992; Blennerhassett \& Bassom 1994). In this type of asymptotic theory, the mean flow varies by a finite amount from the base flow and is therefore called a strongly nonlinear theory. However, the scaling of momentum and heat transport of the nonlinear state is not so different from that of the basic flow, and in this sense, the character of fully developed turbulence is not well captured. Attempts have been made to construct asymptotic solutions with larger amplitudes, but a complete understanding is still lacking. This paper proposes a solution to this long-standing problem.

In the next section, we begin by formulating our problem. Comparisons of the Taylor vortex solutions with previous experiments and simulations are then carried out in section 3. In section 4, a matched asymptotic expansion analysis is performed for the case of Taylor vortices with an aspect ratio about unity. We will see in section 5 that the asymptotic structure of the solution dramatically changes when the axial period becomes asymptotically short. In section 6, how the short-period vortices develop from the laminar solution is investigated in detail using a matched asymptotic expansion. Finally, in section 7, the main findings are summarised and discussed.

\section{Formulation of the problem}

Taylor-Couette flow can be described by the Navier-Stokes equations in the cylindrical coordinates $(r,\theta,z)$.
If the flow is axisymmetric, the governing equations are written as
%\textcolor{black}{change to vector form}
\begin{subequations}\label{NSeq}
\begin{eqnarray}
Du-r^{-1}v^2=-\partial_r p+\triangle u -r^{-2}u,\label{moma}\\
Dv+r^{-1}uv=\triangle v -r^{-2}v,\\
Dw=-\partial_z p+\triangle w,\label{momc}\\
r^{-1}\partial_r(ru)+\partial_z w=0.
%
%Du-r^{-1}v^2=-\partial_r p+\triangle u -r^{-2}(u+2\partial_{\theta}v),\label{moma}\\
%Dv+r^{-1}uv=-r^{-1}\partial_{\theta}p+\triangle v -r^{-2}(v-2\partial_{\theta}u),\\
%Dw=-\partial_z p+\triangle w,\label{momc}\\
%r^{-1}\partial_r(ru)+r^{-1}\partial_{\theta}v+\partial_z w=0.
\end{eqnarray}
\end{subequations}
The operators $D$ and $\triangle$ are defined as 
$D=\partial_t+u\partial_r+w\partial_z$ and 
$\triangle =\partial_r^2+r^{-1}\partial_r+\partial_z^2.$ 
The length and velocity scales are chosen so that the cylinder gap is unity and the no-slip conditions on the cylinder walls are described as
\begin{eqnarray}
(u,v,w)=(0,R_o,0)\qquad \text{at} \qquad r=r_o,\\
(u,v,w)=(0,R_i,0)\qquad \text{at} \qquad r=r_i,
\end{eqnarray}
using the Reynolds numbers associated with the rotation of the inner and outer cylinders, $R_i$ and $R_o$.
Note that our non-dimensionalisation implies that using the radius ratio $\eta=r_i/r_o < 1$ the inner and outer radii are specified as
\begin{eqnarray}
r_i=\frac{\eta}{1-\eta}, \qquad r_o=\frac{1}{1-\eta},
\end{eqnarray}
respectively.
The circular Couette flow solution is written as
\begin{eqnarray}
(u,v,w)=(0,v_b,0),\qquad v_b(r)=\frac{R_o-\eta R_i}{1+\eta}r+\frac{\eta^{-1} R_i-R_o}{1+\eta}\frac{r_i^2}{r}. \label{base}
\end{eqnarray}

For other nontrivial solutions of (\ref{NSeq}) periodicity is imposed in the interval %$\theta \in [0,2\pi/m]$ and 
$ z\in [0,2\pi/k]$ using the axial wavenumber $k$. %The axial mass flux is set to be zero. 
%where $L f = f_{rr}+r^{-1}f_r-r^{-2}f +f_{zz} =(r^{-1}(rf)_r)_r+f_{zz} =r^{-2}(r^3(r^{-1}f)_r)_r+f_{zz} $.
The momentum equations (\ref{moma})-(\ref{momc}) can be simplified as
\begin{subequations}\label{axieqs}
\begin{eqnarray}
r^{-1} D(rv) = \triangle v -\frac{v}{r^2},\qquad rD\left ( \frac{\omega}{r} \right )=\triangle \omega -\frac{\omega}{r^2}+\frac{\partial_z v^2}{r},
%
%r^{-2} \{ \Psi_r (rv)_z - \Psi_z (rv)_r \} = v_{rr}+r^{-1}v_r-r^{-2}v +v_{zz},\\
%\Psi_r \left ( \frac{\omega}{r} \right )_z - \Psi_z \left ( \frac{\omega}{r} \right )_r = {\omega}_{rr}+r^{-1}{\omega}_r-r^{-2}\omega +\omega_{zz} +\frac{2v v_z}{r},
\end{eqnarray}
using the azimuthal vorticity 
\begin{eqnarray}
\omega=\partial_z u-\partial_r w=-\{\partial_r (r^{-1}\partial_r \Psi)+r^{-1}\partial_z^2\Psi \}
\end{eqnarray}
and the Stokes streamfunction $\Psi$. The roll-cell velocity can be reconstructed as $u=-r^{-1}\partial_z \Psi$ and $w=r^{-1}\partial_r \Psi$.
\end{subequations}
Steady solutions of the above system can be computed without regarding their stability by using the numerical code used in \textcolor{black}{our} previous studies (Deguchi \& Altmeyer 2013; Deguchi et al. 2014). The code is based on the Newton-Raphson method applied \textcolor{black}{to} the Chebyshev-Fourier discretised system. To calculate the Taylor vortex with an aspect ratio of about unity, we typically used up to 250th Chebyshev polynomials and 250th Fourier \textcolor{black}{harmonics}. This spatial resolution is more than sufficient for $Ta=O(10^{10})$. We shall see that when $k$ is large, we can reach higher Taylor numbers. In this case, the highest degree of Chebyshev polynomials was increased to 450 to fully resolve the very thin near-wall boundary layer structures.

Instead of the two Reynolds numbers, \textcolor{black}{the majority of} high Taylor number TCF studies summarised in Grossmann et al. (2016) used the Taylor number $Ta$ and the rotation rate $a$. They are easily found by the standard parameters as
\begin{eqnarray}
Ta^{1/2}=\frac{(1+\eta)^3}{8\eta^2}(R_i-\eta R_o), \qquad a=-\eta \frac{R_o}{R_i}.
\end{eqnarray}
\textcolor{black}{
The former parameter is similar to the shear Reynolds number used in Dubrulle et al. (2005) and is zero for the rigid body rotation case (their second parameter, the rotation number, is a function of $a$ and $\eta$).
}
The Nusselt number is defined by
\begin{eqnarray}
Nu=\left . \frac{\partial_r (r^{-1}\overline{v})}{\partial_r (r^{-1}v_b)}\right |_{r=r_i}=\left . \frac{\partial_r (r^{-1}\overline{v})}{\partial_r (r^{-1}v_b)}\right |_{r=r_o},\label{Nusselt}
\end{eqnarray}
where 
\textcolor{black}{
\begin{eqnarray}
\overline{v}(r)=\frac{k}{2\pi}\int^{2\pi/k}_0  v(r,z) dz
\end{eqnarray}
}
is the mean azimuthal velocity.

In the limit $\eta \rightarrow 1$, the gap becomes much narrower than the cylinder radius, and the local flow can be represented in Cartesian coordinates. As noted by Deguchi (2016) there are several variations on the narrow gap limit, two of which are relevant to this paper. One is of course the rotating plane Couette flow, where the system is in perfect agreement with RBC if the Prandtl number is unity (see \textcolor{black}{Chandrasekhar 1961, Robinson 1967, Veronis 1970, and Lezius \& Johnston 1976,} Brauckmann \& Eckhardt 2017, for example). The other utilises an argument similar to the derivation of the G\"ortler vortex (Hall 1983), and is used, for example, by Denier (1992). For completeness, the difference between the two limits is highlighted in Appendix A.

%\textcolor{black}{Inverse Rossby number,  although the term centrifugal instability is also used.}

\section{Comparison of the steady solutions and the experiments}

\begin{figure}
\begin{center}
  \includegraphics[width=0.8\textwidth]{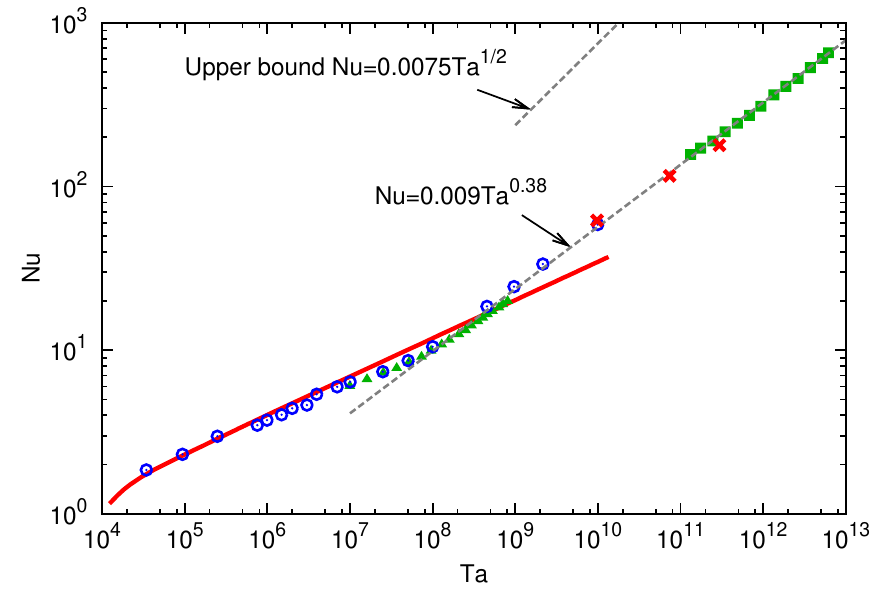}
\end{center}
\caption{
The variation of Nusselt number $Nu$ with respect to Taylor number $Ta$. The outer cylinder is fixed ($a=0$). $\eta=5/7\approx 0.714$. 
The red solid curve is the Taylor-vortex solution branch with the fixed wavenumber $k=3$. 
The red crosses are the same solutions, but the wavenumber is optimised to maximise $Nu$.
The blue circles are the three-dimensional direct numerical simulations by Ostilla et al. (2013) and Ostilla-M\'onico (2014a).
The cyan triangles and squares are the experiments by Lewis \& Swinney (1999) and 
%$\eta=0.724$.
Dennis et al. (2011), respectively. 
The simulation and experimental results are time-averaged data.
\textcolor{black}{The best theoretical upper bound known to date has the asymptotic form $Nu=0.0075\,Ta^{1/2}$ according to Ding \& Marensi (2019). }
%$\eta=0.716$.
%$\eta=5/7$.
%
%The numerical and theoretical results are for 
%Slightly different radius ratios were used in the experiments ($\eta=0.724$ for Lewis \& Swinney (1999), $\eta=0.716$ for Dennis et al. (2011)).
}
\label{fig}
\end{figure}

\textcolor{black}{
Here we use the radius ratio $\eta=5/7$ and fix the outer cylinder ($a=0$) to compute the Taylor vortex solutions. Significant deviations from the narrow gap approximation can be observed at this radius ratio. That parameter choice is frequently used in experiments and numerical simulations (see section 3 of Grossman et al. 2016) and is hence convenient for the comparison. 
In experiments, it has been confirmed that the effect of endwalls on torque/mean flow measurements is negligible (see van Gils et al. (2012), for example).
}

\textcolor{black}{
The red curve in figure 1 shows the Taylor vortex solution calculated with the fixed axial wavenumber $k$.
%the axial wavenumber fixed at $k=3$. 
According to the simulations, the Taylor cells are relatively robust even in the numerically generated classical turbulent flows, with a cell aspect ratio of approximately unity. More specifically, direct numerical simulations (DNS) by Ostilla et al. (2013) imposed the axial periodicity $2\pi$ and typically observed three vortex pairs (i.e. $k=3$ modes). This motivated the choice of $k=3$ for our Taylor vortex computation. 
}

The solution branch bifurcates from the circular Couette flow at $Ta\approx 10^4$, and as the Taylor number increases, it loses stability with respect to three-dimensional perturbations. The onset of turbulence is $Ta\approx 3\times 10^6$. The blue circles in the figure are the DNS by Ostilla et al. (2013) and Ostilla-M\'onico et al. (2014a). 
\textcolor{black}{Despite the relatively short axial periodicity imposed, the simulations are known to agree with the experimental results (the squares and triangles in figure 1). Both the experiments and simulations} indicate the existence of a transition point $Ta\approx 10^8$ at which the behaviour of $Nu$ changes. The turbulence below/above the transition point is referred to as the classical/ultimate regime. The Nusselt number of the ultimate turbulence has the scaling $Nu\propto Ta^{\beta}$ with the exponent $\beta\approx 0.38$, which is also seen in the RBC experiments (He et al. 2012). The empirical asymptotic prediction \textcolor{black}{$Nu=0.009\,Ta^{0.38}$} sits well below the theoretical upper bound of $Nu$ derived by Ding \& Marensi (2019). The exponent 1/2 is typical for upper bounds using the energy method. Similar asymptotic behaviour of $Nu$ but with a logarithmic correction has been proposed using the phenomenology of the log law of turbulent boundary layers (see Grossmann et al. 2016). It is, however not known whether such scaling can be clearly observed in experiments.

\begin{figure}
\centering
\begin{tabular}
[c]{cc}
(a)\hspace{10mm}& (b)\hspace{10mm}\\
\includegraphics[scale=0.9]{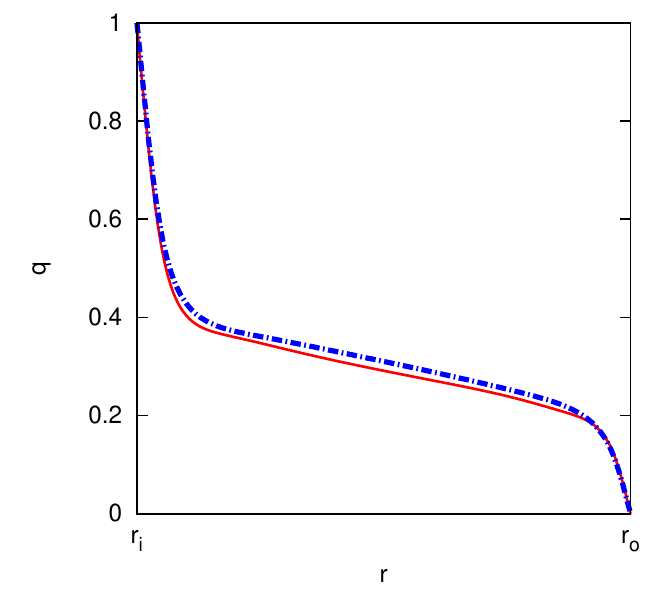} &
\includegraphics[scale=0.9]{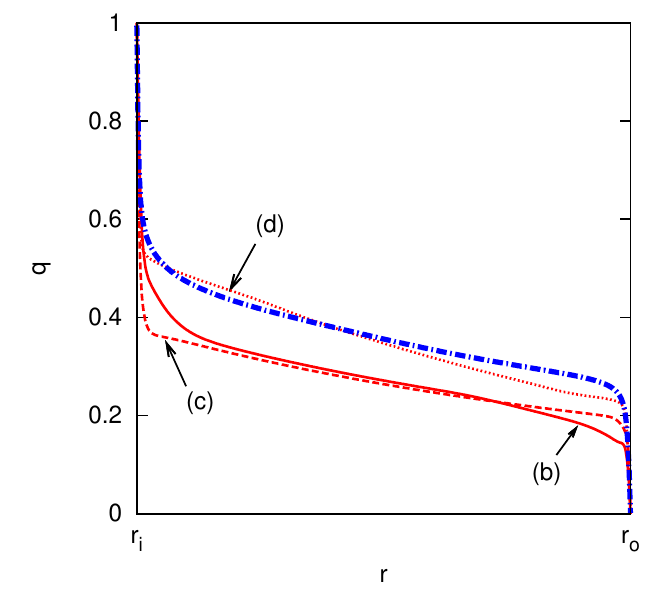}
\end{tabular}
\caption{
The mean angular velocity \textcolor{black}{$q$ defined in (\ref{defq})} for $\eta=5/7, a=0$.
(a) The classical turbulence regime $Ta=9.52\times 10^6$.
The red solid curve is the Taylor vortex solution.
The blue dot-dashed curve is the time-averaged DNS result by Ostilla et al. (2013).
(b) The ultimate turbulence regime. The blue dot-dashed curve is the time-averaged DNS result by Ostilla-M\'onico et al. (2014b) ($Ta=10^{10}$).
The other curves are the Taylor vortex solutions shown in figure 3 (b)-(d) ($Ta=9.75\times 10^9$). 
}
\label{fig}
\end{figure}

\begin{figure}
\hspace{20mm} (a)\\
\begin{center}
  \includegraphics[width=0.8\textwidth]{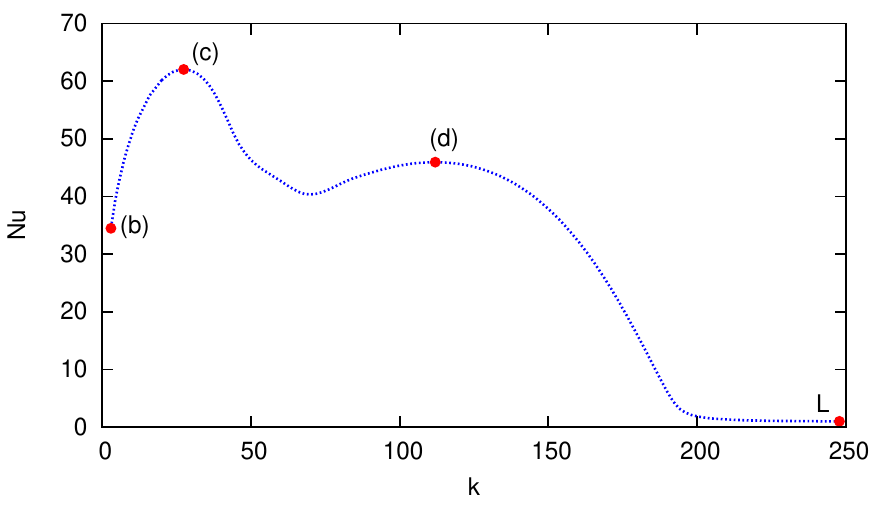}\\
\hspace{0mm}(b) $k=3$\hspace{20mm}(c) $k=27.4$\hspace{20mm}(d) $k=112$ \hspace{40mm}
  \includegraphics[width=0.9\textwidth]{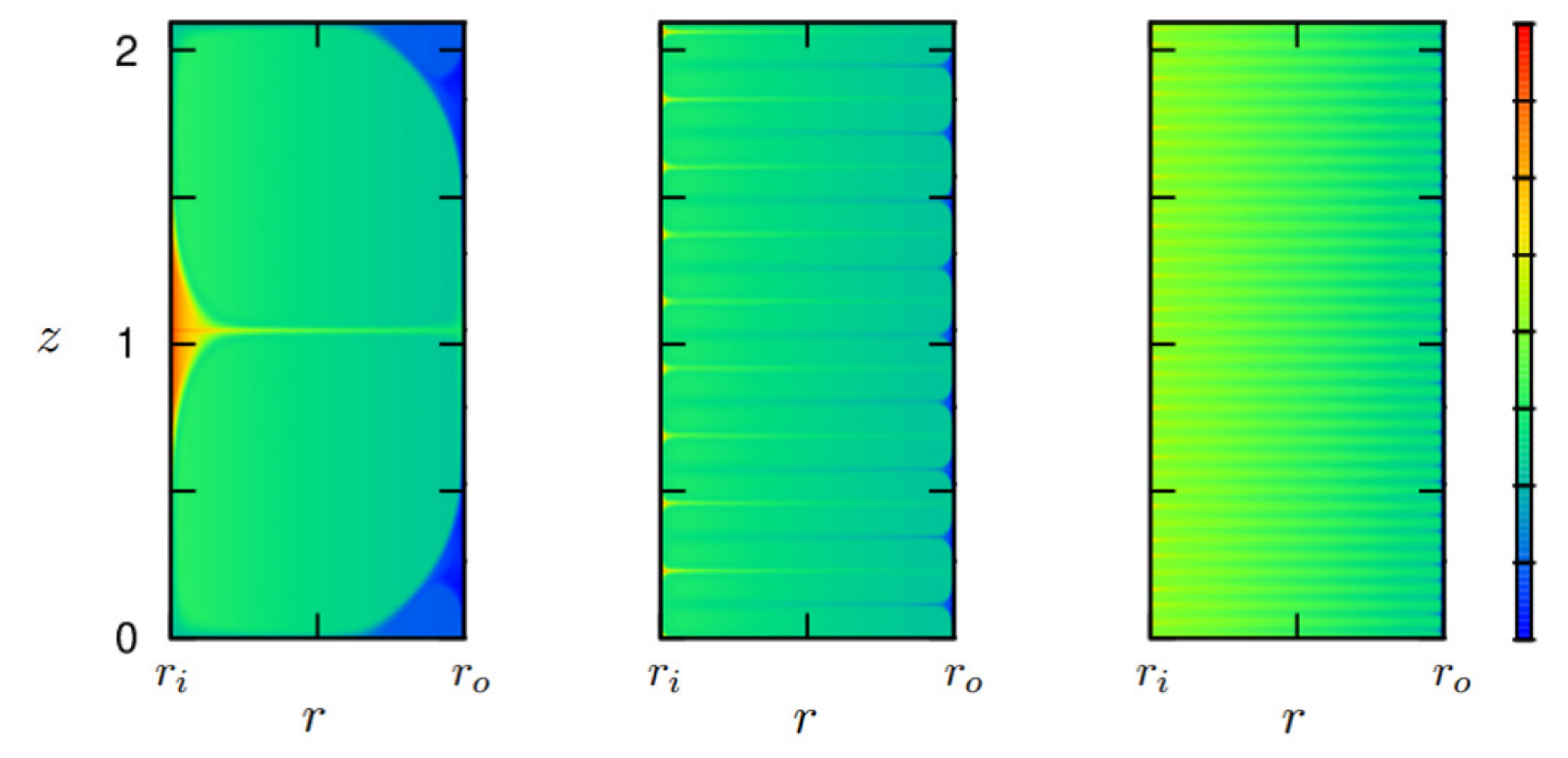}\\
    \includegraphics[width=0.7\textwidth]{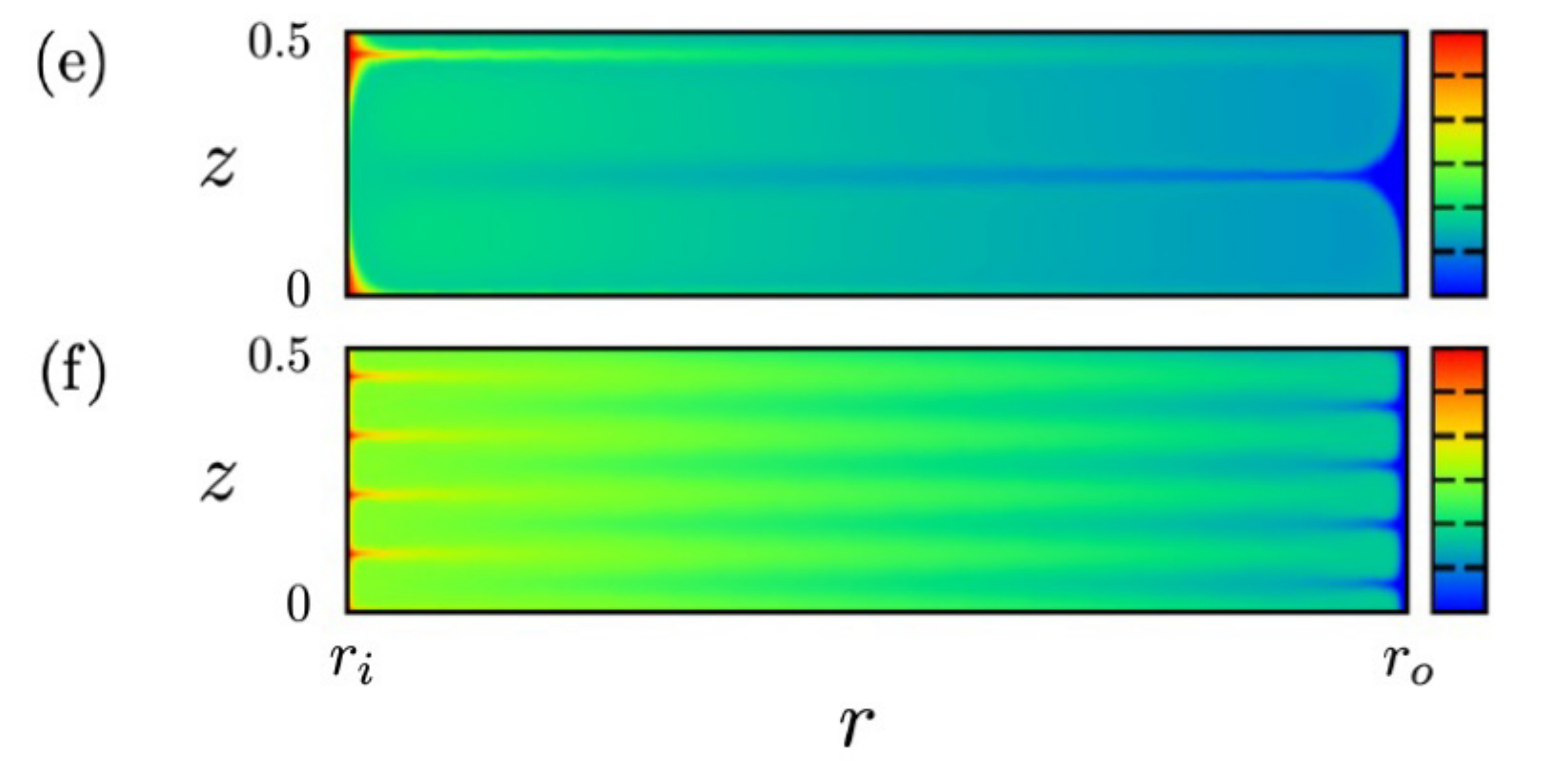}
\end{center}
\caption{
Axial wavenumber dependence of the Taylor vortex solutions. The parameters are $\eta=5/7, R_i=8\times 10^4, R_o=0$, which correspond to $Ta=9.75\times 10^9$ in figure 1. 
(a) 
The bifurcation diagram. The blue dotted curve is the Taylor vortex solution branch. This branch bifurcates from the linear critical point of the circular Couette flow (`L' in the figure). \textcolor{black}{There is another linear critical point at very small $k$, but computing the bifurcating solution branch at this Taylor number is difficult and hence it is omitted.}
(b)-(d) Azimuthal velocity $v$ at the selected points in figure 3a.
%(b) $k=3$, (c) $k=27.4$, (d) $k=112.$ 
The colour bar range is [0,80000]. \textcolor{black}{All the solutions posses the reflection symmetry in $z$.
(e) and (f) are enlarged views of parts of (c) and (d), respectively. The colour bar range is changed to [10000,70000] in the enlarged figures.}
%For relatively small $k$ up to 250th Chebyshev polynomials and 250th Fourier modes are used.
%For high wavenumbers more than 400th Chebyshev polynomials are needed to fully resolve the solutions.
%CB range: [0:40000] (need to change scaling).
}
\label{fig}
\end{figure}

As long as the solution has an aspect ratio of about unity, it captures the nature of classical turbulence surprisingly well. This is best illustrated by a comparison of mean flows shown in figure 2a.
Here
\begin{eqnarray}
q=\frac{\overline{v}/r-R_o/r_o}{R_i/r_i-R_o/r_o} \in [0,1] \label{defq}
\end{eqnarray}
is the shifted and normalised mean angular velocity (denoted by $\langle \overline{\omega}\rangle_z$ in Ostilla et al. 2013).
At sufficiently high Taylor numbers, the boundary layer formation near the cylinder walls is clearly visible. As seen in figure 1, the solution gives a reasonable estimate of $Nu$ in the classical turbulence regime. The cross-section of the Taylor vortex (see figure 3b) reveals that the boundary layer actually surrounds the vortex core, where the azimuthal velocity appears to be rather uniform. The dynamics of the boundary layers in the classical turbulence are known to be somewhat quiescent (see Ostilla et al. 2013, for example), and their qualitative structure is reminiscent of figure 2b. The literature summarised in Grossman et al. (2016) often \textcolor{black}{implies} that the Prandtl-Blasius theory could be applied to the boundary layer, and we shall see in section 4 that this is, in fact, true for the asymptotic limit of the steady solution. 
%$Re_w$
%to characterise the nature of the core roll flow, 
%the wind Reynolds number $Re_w$ is often measured.
%i.e. the non-dimensional roll cell speed in the core region of the vortex 
Moreover, the theory to be presented in section 4 yields the scaling of $Nu \propto Ta^{0.25}$ and $Re_w \propto Ta^{0.5}$, which are well agreeing with the turbulent observations.
 \textcolor{black}{
 Here $Re_w$ is  the wind Reynolds number, i.e. the typical roll cell circulation speed normalised by the viscous velocity scale $\nu/d$ ($\nu$ is the kinematic viscosity of the fluid and $d$ is the cylinder gap).
The precise definition of the wind Reynolds number varies according to the literature. 
Huisman et al. (2012) measured the standard deviation of the radial velocity $u$, while Ostilla et al. (2013) computed the average of $u^2+w^2$ and then square rooted it. 
Both definitions give the same scaling, but with different prefactors.
}

In the ultimate turbulence, on the other hand, the situation appears to be more intricate. The snapshots of turbulence are typically accompanied by eddies of a vast scale, whereby the large-scale Taylor vortex is blurred in the time-averaged field. In particular, small turbulent eddies appearing in the boundary layer have been pointed out as a critical qualitative difference between the two turbulent regimes separated by the transition point seen in figure 1 (Dong 2007, Ostilla-M\'onico et al. 2014a). Therefore solutions with larger wavenumbers may play a more critical role in the turbulent dynamics. Our Nusselt number results also support this speculation. As seen in figure 1, the Taylor vortex with fixed $k$ underestimates experimental $Nu$ in the ultimate regime. However, if $k$ is optimised to maximise $Nu$ (the crosses in figure 1), it can \textcolor{black}{reach} the experimental values for $Ta \lesssim 10^{11}$. 

Figure 3a shows how $Nu$ changes as the wavenumber $k$ of the Taylor vortex is varied. The $Nu$ curve has two local maxima. \textcolor{black}{The bimodal distribution of $Nu$ and the scaling of the two peaks are remarkably} similar to those seen in the RBC roll cell computation by Waleffe et al. (2015) and Sondak et al. (2015). Based on their calculations at several Rayleigh numbers $Ra$, the first (second) peak has the wavenumber scaling $k\propto Ra^{0.217}$ ($Ra^{0.256}$) with the Nusselt number scaling exponent $\beta$ slightly larger (smaller) than 0.31. We shall provide a theoretical rationale for the scaling in sections 5 and 6.

For the ultimate turbulence regime, the approximation of the turbulent mean flow by the $k=3$ state is slightly worse (figure 2b). 
\textcolor{black}{
However, the mean flow of course varies with $k$. 
%As remarked earlier, the solutions with fixed aspect ratio of about unity is no more representative of turbulence at this $Ta$.
As the value of $k$ is increased from 3, the mean flow of the solution in the core region %does not change much up to the first peak, but then 
approaches a turbulent profile in the vicinity of the second peak.}
%However, in the core region, the extent of this error is about the range of change that occurs when $k$ is varied. 
\textcolor{black}{Figures 3c,d show} the flow field at each peak, which is also examined in detail in the following two sections.

\section{Taylor vortices with an $O(1)$ cell aspect ratio}

In figure 2a, we saw that some kind of homogenisation occurs in the core region of the Taylor vortex, and a thin boundary layer emerges around it. Such a flow structure looks very much like the high Rayleigh number RBC roll cell, which was studied intensively from \textcolor{black}{the 1950's to the 1970's,} for example, by Pillow (1952), Robinson (1967), and Wesseling (1969). More recently, it was reported that when the slip walls are \textcolor{black}{imposed,} the asymptotic solution can be \textcolor{black}{calculated} semi-analytically and is in good agreement with the numerical solutions (Chini \& Cox 2009; Hepworth 2014). However, as noted by Robinson (1967), for no-slip walls, the scaling and structure of the boundary layer have to be modified, which makes analytical progress difficult. The situation in the Taylor vortex is close to the latter problem, as we shall see below.

First, let us examine the core region. For RBC, the roll cell vorticity in this region is known to become constant due to the Prandtl-Batchelor theorem, which states that the homogenisation of the vorticity occurs in the region where the streamlines are closed in two-dimensional inviscid flows (Prandtl 1904; Feynman \& Lagerstrom 1956; Batchelor 1956). Moreover, the temperature becomes constant as well because the temperature equation has \textcolor{black}{a similar structure to that of} the vorticity equation (see Moore \& Weiss 1973, for example). To better see what precise physical quantities are homogenised in the core of the Taylor vortex, \textcolor{black}{it is desirable} to choose large gaps. Figure 4 shows the results for $\eta=0.5$. The colour map shown in figure 4b clearly indicates that it is actually angular momentum $rv$ that becomes uniform in the region where the streamlines are closed. Also, figure 4c shows that it is not the azimuthal vorticity $\omega$ that homogenises, but $\omega/r$.
\begin{figure}
\hspace{20mm}(a)\hspace{30mm}(b)\hspace{30mm}(c)\hspace{40mm}
\begin{center}
  \includegraphics[width=0.9\textwidth]{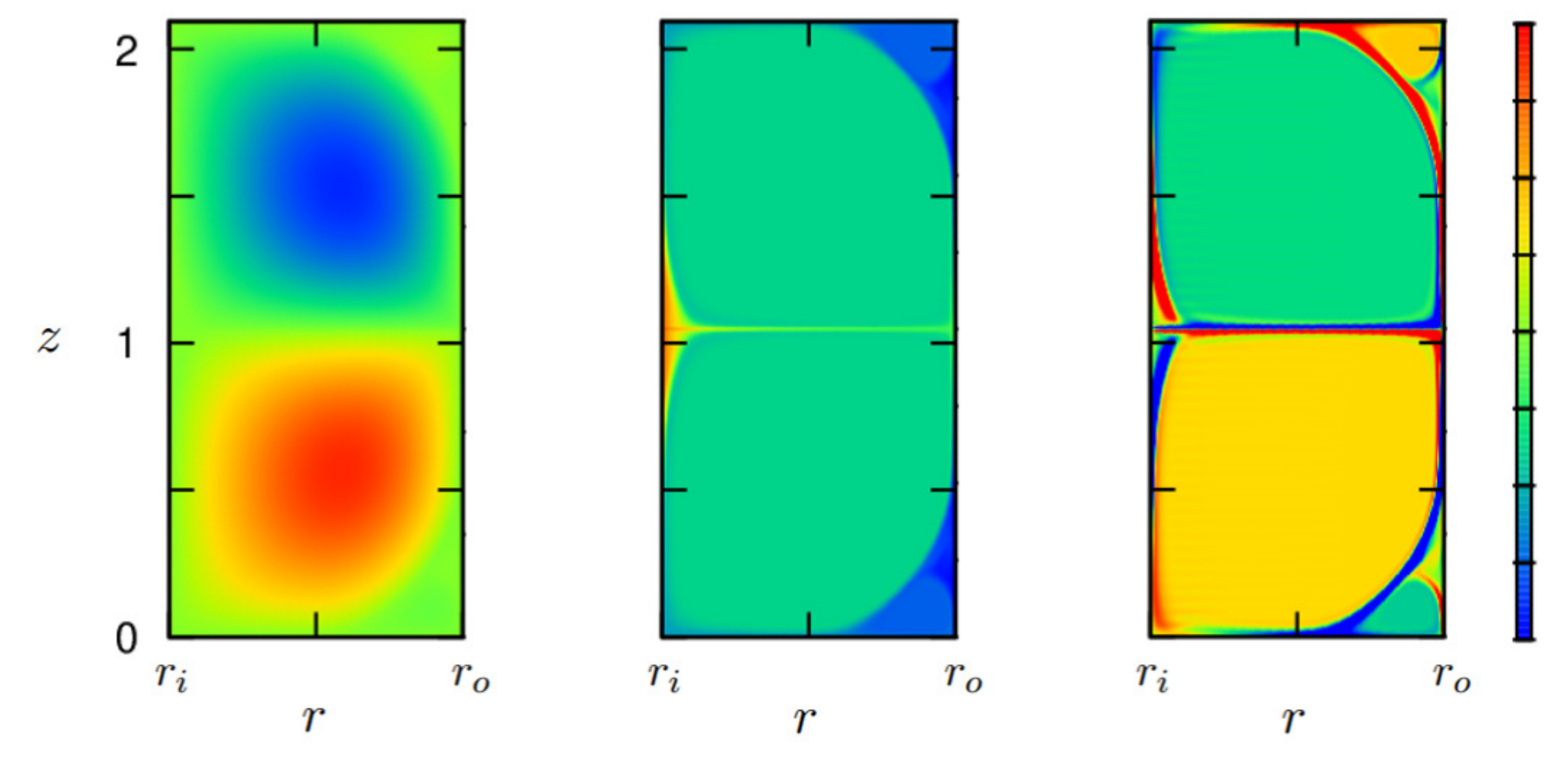}
\end{center}
\caption{The flow field of the Taylor vortex for $\eta=0.5, k=3$. The Reynolds numbers used are $R_i=8\times 10^4, R_o=0.25 R_i$, corresponding to $Ta=1.40\times 10^{10}$, $a=-1/8$.
(a) The Stokes streamfunction $\Psi$. The colour bar range is [-7500,7500].
(b) The angular momentum $rv$. The colour bar range is [40000,80000].
(c) The modified azimuthal vorticity $\omega/r$. The colour bar range is [-120000,120000].
\textcolor{black}{This quantity is uniformly distributed in the core to a value of about $\pm 37$\% of the colour bar.}
%\textcolor{black}{For (a) and (c)the centre of the colour bar is zero.}
%$L=240, M=250.$
%CB range: [-15000:15000], [40000:80000], [-15000:15000] (need to change scaling). 
%CB range: [-15000:15000], [80000:160000], [-60000:60000] (need to change scaling). 
}
\label{fig}
\end{figure}

Building on the above observations, now we introduce the new variables $\Gamma=Ta^{-1/2}rv$ and $\Omega=\omega/r$, whereby the governing equations (\ref{axieqs}) are rewritten as
\begin{subequations}\label{Taeq}
\begin{eqnarray}
 \Psi_r \Gamma_z - \Psi_z \Gamma_r  =(r^3(r^{-2}\Gamma)_r)_r+r\Gamma_{zz},\label{Gammaeq}\\
\Psi_r \Omega_z - \Psi_z \Omega_r = (r^{-1}(r^2\Omega)_r)_r+{r\Omega}_{zz} 
+Ta\frac{2\Gamma \Gamma_z}{r^3},\label{Omegaeq}\\
\Omega=-r^{-1}\{(r^{-1}\Psi_r)_r+r^{-1}\Psi_{zz} \label{Psieq}\}.
\end{eqnarray}
\end{subequations}
Here, the subscript $r$ or $z$ denotes a partial differentiation.
The no-slip conditions become
\begin{subequations}
\begin{eqnarray}
\Psi=\Psi_r=0, \qquad \Gamma=\Gamma_i\equiv \frac{8\eta^3}{(1-\eta)(1+\eta)^3(1+ a)}\qquad \text{at} \qquad r=r_i,\\
\Psi=\Psi_r=0, \qquad \Gamma=\Gamma_o\equiv -\frac{8\eta  a}{(1-\eta)(1+\eta)^3 (1+ a) }\qquad \text{at} \qquad r=r_o.
\end{eqnarray}
\end{subequations}
If $\Gamma$ is considered as temperature, $\Omega$ as roll cell vorticity and $Ta$ as Rayleigh number, one may notice that the structure of the equations is very similar to that of the two-dimensional RBC. 
\textcolor{black}{The last term in the right hand side of (4.1b) corresponds to the Coriolis force when the RPCF limit is taken, and it plays a role of buoyancy. (This term is not exactly the Coriolis force unless the system is in the RPCF limit, but for simplicity we will call it `Coriolis force' hereafter; see Appendix A also.)} \textcolor{black}{This analogy in the sense of the structure of the equations would explain why $\Gamma$ and $\Omega$ become constant in the Taylor vortex core, as seen in figure 4, at least on an intuitive level.}

\begin{figure}
\begin{center}
  \includegraphics[width=0.85\textwidth]{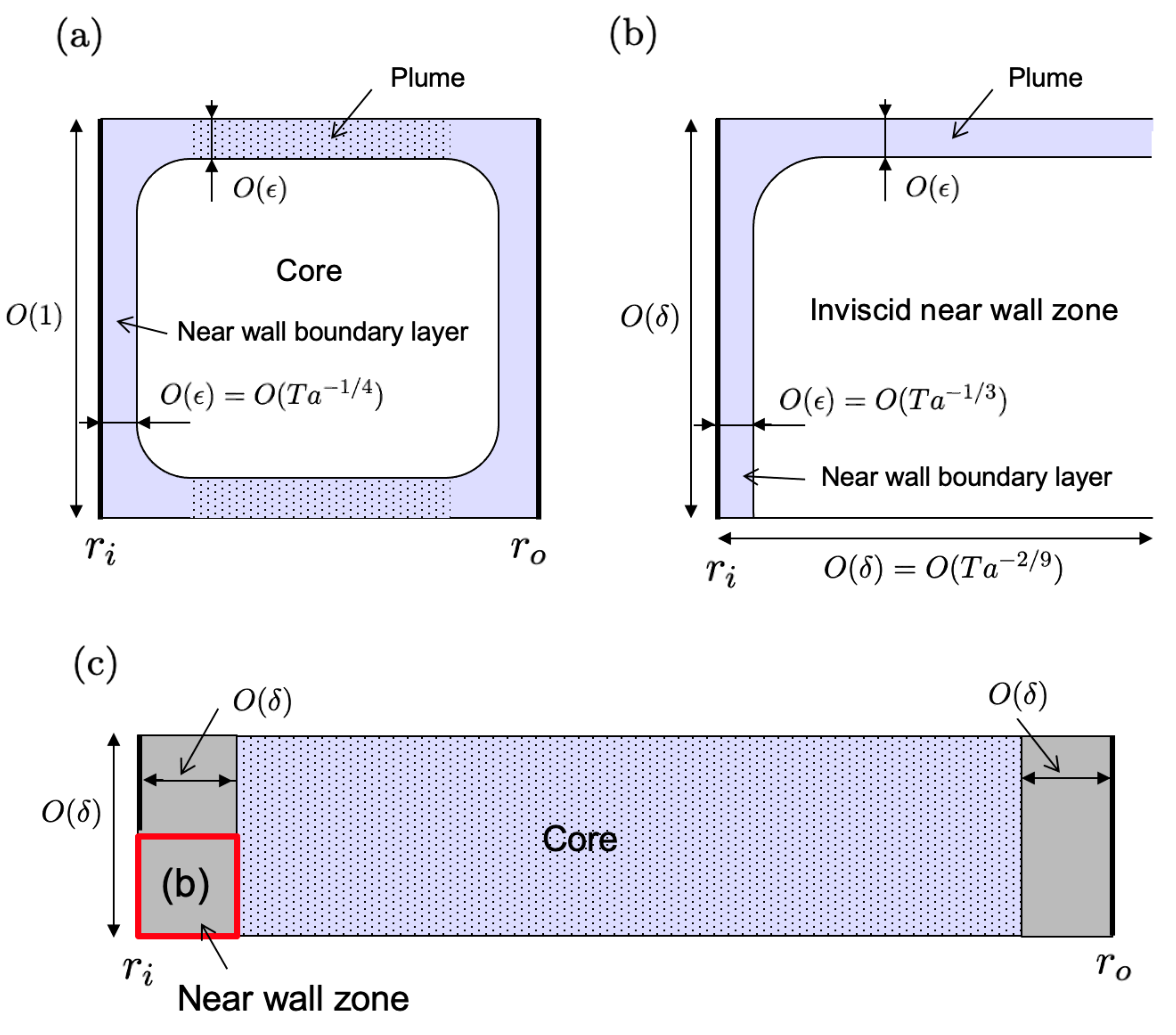}
\end{center}
\caption{
Sketch of the asymptotic states. In the blue shaded region \textcolor{black}{viscosity} is not negligible. In the dotted region Coriolis force is at work. (a) Taylor vortex with aspect ratio of order unity ($k=O(1)$). (b,c) The first peak state ($k=O(Ta^{2/9})$).
(b) is the close up of the near wall zone enclosed by the red lines in (c). 
}
\label{fig}
\end{figure}

In fact, following Batchelor (1956), a mathematical argument can be developed. First, assuming the typical roll cell circulation strength $Re_w$ (the wind Reynolds number)
 is asymptotically large, we expand $\Psi=Re_w\Psi_0+\cdots, \Gamma=\Gamma_0+\cdots, \Omega=Re_w\Omega_0+\cdots$. Then the leading order part of (\ref{Gammaeq}) becomes $\Psi_{0r} \Gamma_{0z}-\Psi_{0z} \Gamma_{0r}=0$, which suggests that the function $\Gamma_0$ only depends on $\Psi_0$. Now, in the $r$-$z$ plane, take a region $\mathcal{A}$ enclosed by a streamline $\mathcal{C}$ oriented counterclockwise (thus $\Psi$ is constant along $\mathcal{C}$).
%Stokes theorem
%\begin{eqnarray}
%\int_A (\nabla \times \mathbf{F}) \cdot (-\mathbf{e}_{\theta}) dr dz =\oint_C \mathbf{F}\cdot d\mathbf{l}.
%\end{eqnarray}
Integrating (\ref{Gammaeq}) over $\mathcal{A}$, we obtain
\begin{eqnarray}
\int_{\mathcal{A}} \{\Psi_r \Gamma_z - \Psi_z \Gamma_r\} drdz =\int_{\mathcal{A}}  \{(r^3(r^{-2}\Gamma)_r)_r+(r^3(r^{-2}\Gamma)_z)_z\} drdz.\label{Gammaint}
\end{eqnarray}
\textcolor{black}{
The left hand side vanishes because upon using the Stokes theorem it becomes
\begin{eqnarray}
\int_{\mathcal{A}} \{(\Psi_r \Gamma)_z - (\Psi_z \Gamma)_r\} drdz
=-\oint_{\mathcal{C}}  \Gamma \{\Psi_z \mathbf{e}_z+\Psi_r \mathbf{e}_r\} \cdot d\mathbf{l}=0.
\end{eqnarray}
The second equality comes from the fact that the gradient of $\Psi$ and the line element on $\mathcal{C}$, $d\mathbf{l}$, are orthogonal on the streamline.}
While the leading order part of the right hand side can be transformed by \textcolor{black}{again applying} the Stokes theorem:
\begin{eqnarray}
0&=&\oint_{\mathcal{C}}  r^3\{(r^{-2}\Gamma_0)_r\mathbf{e}_z-(r^{-2}\Gamma_0)_z\mathbf{e}_r\} \cdot d\mathbf{l}\nonumber \\
&=&\frac{d\Gamma_{0}}{d\Psi_0}\oint_{\mathcal{C}}  r^2\{r^{-1}\Psi_{0r}\mathbf{e}_z-r^{-1}\Psi_{0z}\mathbf{e}_r\} \cdot d\mathbf{l}
-2 \Gamma_0 \oint_{\mathcal{C}}  \mathbf{e}_z \cdot d\mathbf{l} \nonumber \\
&=&\frac{d\Gamma_{0}}{d\Psi_0}\oint_{\mathcal{C}} r^2\{u_0\mathbf{e}_r+w_0\mathbf{e}_z\}\cdot d\mathbf{l}.
\end{eqnarray}
Here $(u_0, w_0)=(-r^{-1}\Psi_{0z},r^{-1}\Psi_{0r})$ is the leading order part of the roll velocity. The integral in the last line should not vanish because we are assuming the existence of a strong circulation due to the swirling motion of the Taylor roll. Thus the equation implies $\frac{d\Gamma_{0}}{d\Psi_0}=0$ at the value of $\Psi_0$ on ${\mathcal{C}}$ we choose. The above argument holds for any region enclosed by a streamline, and hence the value of $\Gamma_0$ must be constant in the core region. 
\textcolor{black}{The argument above does not change when an arbitrary constant is added to $\Gamma$.
This implies that if power series asymptotic expansion of $\Gamma$ is considered the homogenisation also occurs in all the higher-order terms as long as they have a spatial scale of $O(1)$.
Therefore for steady solutions, the non-uniform component in $\Gamma$ is exponentially small.}

Likewise we can show that the value of $\Omega_0$ is constant as well in the core. By integrating (\ref{Omegaeq}) over ${\mathcal{A}}$,
\begin{eqnarray}
\int_\mathcal{A} \{\Psi_r \Omega_z - \Psi_z \Omega_r \} drdz &=&\int_\mathcal{A} \{(r^{-1}(r^2\Omega)_r)_r+(r^{-1}(r^2\Omega)_{z})_z\}drdz\nonumber \\
&&+\int_\mathcal{A} Ta\frac{2\Gamma \Gamma_z}{r^3} dr dz,\label{Omegaint}
\end{eqnarray}
and of course the left hand side should vanish. Since we already know that $\Gamma_0$ is a constant plus an \textcolor{black}{exponentially} small fluctuation, the last term in the right hand side of (\ref{Omegaint}) can be neglected. From (\ref{Omegaeq}) we see that $\Omega_0$ is a function of $\Psi_0$, and thus
the leading order part of integral (\ref{Omegaint})
\begin{eqnarray}
0&=&\oint_{\mathcal{C}} r^{-1}\{(r^2\Omega_0)_r\mathbf{e}_z-(r^2\Omega_0)_{z}\mathbf{e}_r)\}\cdot d\mathbf{l}\nonumber \\
&=&\frac{d\Omega_{0}}{d\Psi_0} \oint_{\mathcal{C}} r^2\{r^{-1}\Psi_{0r} \mathbf{e}_z-r^{-1}\Psi_{0z}\mathbf{e}_r\}\cdot d\mathbf{l}+2\Omega_0 \oint_{\mathcal{C}} \mathbf{e}_z \cdot d\mathbf{l}\nonumber \\
&=&\frac{d\Omega_{0}}{d\Psi_0}\oint_{\mathcal{C}} r^2\{u_0\mathbf{e}_r+w_0\mathbf{e}_z\}\cdot d\mathbf{l},
\end{eqnarray}
yields $\frac{d\Omega_{0}}{d\Psi_0}=0$.

In the above argument, we have assumed that the swirling speed of the rolls is sufficiently large, but to see exactly how large it is, we need to analyse the boundary layer. Let us now take the region $\mathcal{A}$ as large as possible and assume that a viscous boundary layer appears around its boundary $\mathcal{C}$. In this core region (see figure 5a), we have the estimation $Re_w=O(\Psi|_c)= O(\Omega|_c)$, where the subscript $c$ indicates that the physical quantities are measured in the core. At the roll cell perimeter $\mathcal{C}$ without loss of generality, we can set $\Psi=0$. Hence, if the thickness of the boundary layer is written as $\epsilon $, the size of the streamfunction there can be estimated as $O(\Psi|_b)= O(\epsilon \Psi|_c)=O(\epsilon Re_w)$ (the subscript $b$ stands for boundary layer). In order to ensure the viscous-convective balance of (\ref{Omegaeq}) in this thin layer we further require $O(\Psi|_b)=O(\epsilon^{-1})$, and therefore $Re_w=O(\epsilon^{-2})$.

As seen in figure 4c, parts of the boundary layer are attached to the walls, where the flow has to fulfil the no-slip conditions. In order for the streamfunction to be modified in the near-wall boundary layer, both sides of (\ref{Psieq}) must balance, so $O(\Omega|_b)=O(\epsilon^{-1}Re_w)$ using $\partial_r=O(\epsilon^{-1})$ and $O(\Psi|_b)=O(\epsilon^{-1})$. Another essential part of the boundary layer is the plume, where the layer is detached from the wall. When the boundary layer leaves the wall, it typically forms a sharp corner. Similar to the RBC cases, as we round the corner, the sizes of $\Omega$ and $\Gamma$ are unchanged. This can be justified by considering a streamline within the boundary layer. The streamline passes through the $O(\epsilon^{1/2})$ neighbourhood of the corner, where the flow is inviscid, and hence $\Omega$ and $\Gamma$ are functions of the streamfunction; see also the RBC literature introduced at the beginning of this section.

The plume layer is no longer parallel to the wall, and the Coriolis force acting there drives the whole Taylor cell. This plume balance allows us to completely fix the flow scaling in terms of $Ta$. Assuming $\partial_r=O(1)$ and $\partial_z=O(\epsilon^{-1})$ in (\ref{Omegaeq}), the viscous-convective terms of $O(\epsilon^{-2}\Omega|_b)=O(\epsilon^{-5})$ \textcolor{black}{counter-balances} the Coriolis term of $O( \epsilon^{-1} Ta)$ when $\epsilon=Ta^{-1/4}$. Here we used the fact that within the near-wall boundary layer, the size of $\Gamma$ is $O(1)$ from the boundary conditions, and the same size must be used in the plume.

The asymptotic structure can be summarised as follows. Within the core region we use the expansions
\begin{eqnarray}
\Psi=\epsilon^{-2}\omega_0\Psi_0(r,z)+\cdots,\qquad
\Gamma=\gamma_0+\cdots,\qquad
\Omega=\epsilon^{-2}\omega_0+\cdots,
\end{eqnarray}
where $\omega_0$ and $\gamma_0$ are constants. 
The expansion is consistent \textcolor{black}{with} the Prandtl-Batchelor structure to leading order, and $\Psi_0$ must be determined by
\begin{eqnarray}
1=-r^{-1}\{(r^{-1}\Psi_{0r})_r+r^{-1}\Psi_{0zz} \}\label{coreeq}
\end{eqnarray}
in the aforementioned region $\mathcal{A}$ with the boundary condition $\Psi_0=0$ on $\mathcal{C}$.

Let $l$ and $n$ be the distance along $\mathcal{C}$ and its inward normal, respectively. Then using the rescaled normal variable $N=\epsilon^{-1} n$, the flow within the boundary layer can be expanded as
\begin{eqnarray}
\Psi=\epsilon^{-1}\Psi^{(b)}(l,N)+\cdots,\qquad
\Gamma=\Gamma^{(b)}(l,N)+\cdots.
\end{eqnarray}
The leading order equations in the boundary layer are
\begin{eqnarray}
 (\Psi_{l}^{(b)} \partial_{N} - \Psi_{N}^{(b)} \partial_{l})\Gamma^{(b)}  &=&r\Gamma_{NN}^{(b)},\label{GammaeqBL}\\
 (\Psi_{l}^{(b)} \partial_{N} - \Psi_{N}^{(b)} \partial_{l})\Psi_{NN}^{(b)} &=&{r\Psi}_{NNNN}^{(b)}
- \frac{2\cos \varphi}{r}\Gamma^{(b)} \Gamma_{N}^{(b)},\label{OmegaeqBL}
\end{eqnarray}
where $\varphi$ is the angle between the $z$-axis and the $n$-axis.

When the boundary layer is in contact with the cylinder wall, the following boundary conditions must be imposed at $N=0$:
\begin{eqnarray}
\Psi^{(b)}=\Psi_{N}^{(b)}=0,\qquad \Gamma^{(b)} =\Gamma_i\qquad \text{at the inner cylinder,}\\
\Psi^{(b)}=\Psi_{N}^{(b)}=0,\qquad \Gamma^{(b)} =\Gamma_o\qquad \text{at the outer cylinder.}
\end{eqnarray}
For the near-wall boundary layer, $\cos \varphi=0$ so (\ref{OmegaeqBL}) is none other than Prandtl's boundary layer equation. While for large $N$, the flow must match the core solution. Let $U(l)$ be the value of $\Psi_{0n}$ on $\mathcal{C}$, which can be found by the core flow problem (\ref{coreeq}). Then as $N\rightarrow \infty$, the boundary layer solution must satisfy
\begin{eqnarray}
\Psi^{(b)}\rightarrow \omega_0 U(l)N,\qquad \Gamma^{(b)} \rightarrow \gamma_0.
\end{eqnarray}
When the boundary layer is detached from the wall, the similar far-field conditions should be applied on both sides $N\rightarrow \pm \infty$.

The theoretical boundary layer thickness $\epsilon=Ta^{-1/4}$ implies the Nusselt number scaling $Nu\propto Ta^{\beta}$ with $\beta=1/4$. We can check if this is consistent with the angular momentum transport. By averaging (\ref{Gammaeq}) with respect to $z$ and further radially integrating from $r_i$ to $r$, the transport balance can be obtained as
\begin{eqnarray}
-\overline{\Psi_z \widetilde{\Gamma}}=r^3(r^{-2}\overline{\Gamma})_r+\frac{16\eta r_i^2}{(1+\eta)^4}Nu.\label{meanGeq}
\end{eqnarray}
Here we used overline to denote the average with respect to $z$ and wrote $\Gamma=\overline{\Gamma}(r)+\widetilde{\Gamma}(r,z)$. Let us consider this balance at a sufficient distance from both walls. In the core region, $\widetilde{\Gamma}$ would be exponentially small because of the Prandtl-Batchelor theorem, so the transport is extremely inefficient. The plume region of thickness $O(\epsilon)$ is hence the main contributor to the left-hand side, which can be estimated as $O(\epsilon (\partial_z \Psi|_b)\widetilde{\Gamma} |_b)=O(\epsilon^{-1})$ using the scaling $\widetilde{\Gamma}|_b=O(1)$, $\Psi|_b=O(\epsilon^{-1})$. This term indeed balances with the second term on the right-hand side.

\begin{figure}
\begin{center}
  \includegraphics[width=0.8\textwidth]{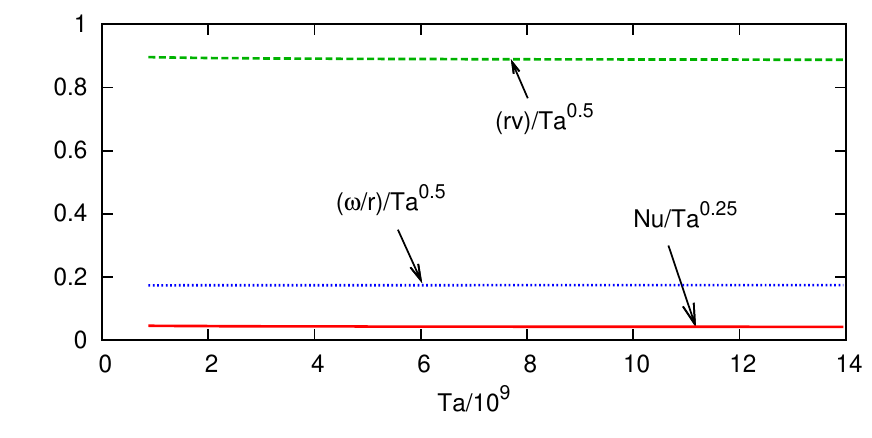}
\end{center}
\caption{
The large $Ta$ asymptotic convergence of the Taylor vortex for $\eta=0.5, k=3, a=-1/8$.
The red solid curve is almost horizontal, implying that the $Nu \propto Ta^{1/4}$ scaling derived for $k=O(1)$ holds.
The green dashed and blue dotted curves are computed by $v$ and $\omega$ measured at the centre of the cell $(r,z)=(r_i+0.5,\pi/2k)$, respectively.
}
\label{fig}
\end{figure}

The asymptotic structure derived here is consistent with the behaviour of the numerical Taylor vortex solution. Figure 6 summarises the numerical results for $\eta=0.5$, $a=-1/8$. The red solid curve in the figure shows the scaling $Nu\propto Ta^{1/4}$, which can also be seen in the data presented in figure 1 ($\eta=5/7, a=0$). The green dashed and the blue dotted curves are $Ta^{-1/2}rv$ and $Ta^{-1/2}\omega/r$ measured at the centre of the roll cell. According to the theory, they converge towards the constants $\gamma_0$ and $\omega_0$, respectively. 
\textcolor{black}{The scaling of $\omega/r$ corresponds to the scaling of $u$, $w$ in the core, and hence the scaling of $Re_w$.
%
%Thus the blue dotted curve implies the scaling $Re_w \propto Ta^{1/2}$, which matches with numerical simulations 
%
%In fact from the Taylor vortex solution, an estimate $Re_w\approx 0.0??\,Ta^{1/2}$ is obtained, which is in good agreement with $Re_w\approx 0.0191\,Ta^{1/2}$ obtained in the DNS by Ostilla et al. (2013).
%
}

%\textcolor{magenta}{
%Moreover, the theoretical asymptotic scaling of the roll cell speed in the core can be used to estimate the Taylor number dependence of the wind Reynolds number $Re_w$.
%Here in turbulent measurements $Re_w$ is usually defined by the root mean square of either or both $u$ and $w$ obtained at a point away from the wall.}

One may have noticed that the exponent $\beta=1/4$ of the Nusselt number differs from the exponent $\beta=1/3$ deduced in the asymptotic analysis of Chini \& Cox (2009) and Hepworth (2014). This discrepancy is not due to the differences in the flow driving mechanism but to the boundary conditions at the walls. If the zero stress condition is imposed, the linear extrapolation of the core streamfunction already satisfies the boundary condition to leading order. This means that the balance $O(\Omega|_b)= O(\epsilon^{-1}Re_w)$ we assumed for the no-slip case is not necessary. For the slip wall case, the magnitude of the vorticity does not change in the core and the boundary layer, so the strong vortex layer seen in figure 4c does not appear. If we use the balance $O(\Omega|_b)=O(\Omega|_c)= O(\epsilon^{-2})$ instead for the scaling argument of the plume we have $\epsilon=Ta^{-1/3}$, as expected. 
 %-----------------

For the slip wall RBC, further analytical progress has been made using the fact that the boundary layer equations become linear and the roll cells are rectangular (Chini \& Cox 2009; Hepworth 2014). In our case, however, we have to rely on numerical calculations because the boundary layer equations are fully nonlinear. Moreover, the shape of the core region is nontrivial due to the small vortices appearing near the corners (see figure 4c). This means that the core and boundary layer equations \textcolor{black}{(4.8), (4.10), (4.11)} need to be iteratively solved by updating the core shapes and constants $\gamma_0$ and $\omega_0$. Such numerical calculations are too challenging and out of the scope of this paper. 
\textcolor{black}{Differences in the structure of the boundary layer also affect the $Pr$ dependence of the asymptotic solution of RBC. For the slip wall case, asymptotic solutions for different $Pr$ can be obtained by rescaling the $Pr=1$ solution. However, this is possible because the boundary layer is linear, which is not the case for the no-slip case.}

\begin{figure}
\centering
\begin{tabular}
[c]{cc}
(a)\hspace{10mm}& (b)\hspace{10mm}\\
\includegraphics[scale=0.9]{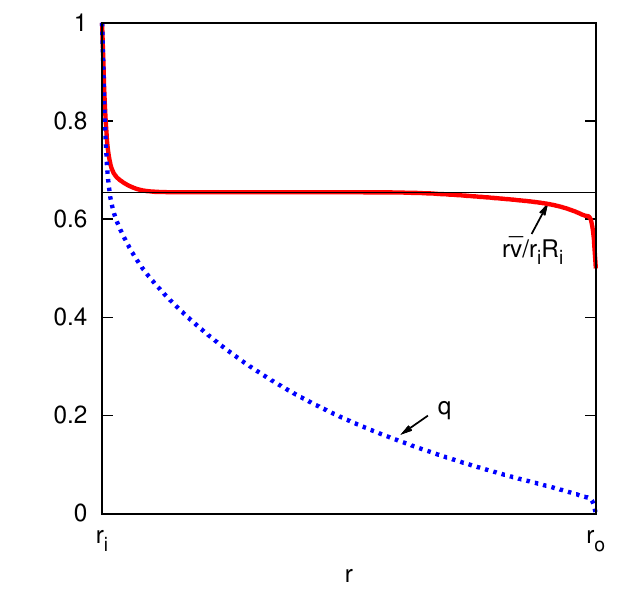} &
\includegraphics[scale=0.9]{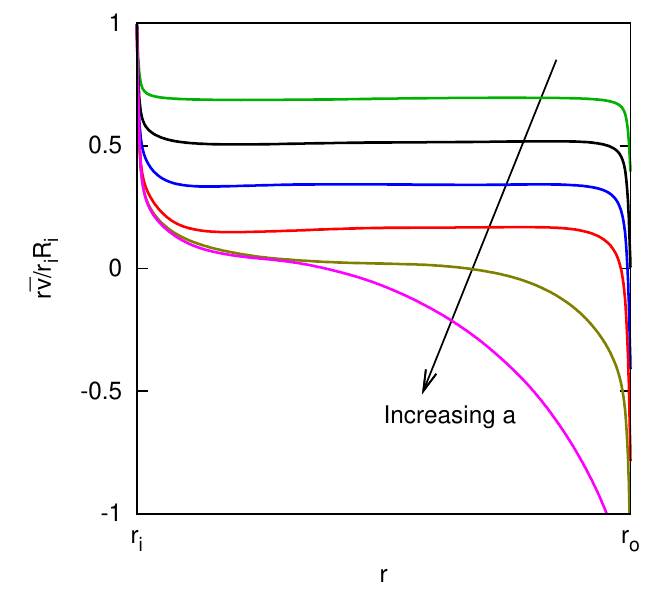}
\end{tabular}
\caption{
Uniform mean angular momentum profiles.
(a) The mean flow of the Taylor vortex solution shown in figure 4 \textcolor{black}{($\eta=0.5, a=-1/8$)}. \textcolor{black}{The red solid curve is the normalised angular momentum, while the blue dotted curve is the mean angular velocity $q$ defined in (\ref{defq})}.
(b) Time average of DNS results for $Ta=10^{10}$, $a=-0.20,0.00,0.21,0.40,0.60,1.00$.
The data are from figure 4 of Ostilla-M\'onico et al. (2014b).
%ultimate turbulence. $Ta=10^{10}$ for DNS. Figure 4 of Ostilla et al. (2014) 
}
\label{fig}
\end{figure}
Finally, we show that the above analysis provides some insights into the structure of the mean flow. As already seen in figure 4b, the angular momentum $rv$ is almost constant in the core region. Therefore, the mean angular momentum $r\overline{v}$ is expected to become a constant away from the wall. This is indeed the case for the numerical Taylor vortex solution (figure 7a). In the studies of TCF turbulence, on the other hand, the mean angular velocity $q$ is usually plotted, and it is often noted that the profile is linear in the core. At first glance, this appears to be the case, for example, when looking at figure \textcolor{black}{2a}, but this is because the cylinder gap \textcolor{black}{($\eta\approx 0.714$)} is too narrow to clearly see the radial dependence of the profile (see figure 7a for the $q$ profile for a wide gap case \textcolor{black}{$\eta=0.5$}). If the numerical results by Ostilla-M\'onico et al. (2014b) are summarised in terms of $rv$, as shown in figure 7b, they clearly show the constant angular momentum property in the core. Note that when $a$ becomes too large, the Taylor vortex appears to favour the vicinity of the inner cylinder, and the homogenisation is only observed there. In the long history of TCF studies, some researchers have also pointed out that the turbulent mean flows might have \textcolor{black}{a} uniform angular momentum zone (Wattendorf 1935, Taylor 1935, Smith \& Townsend 1982, Lewis \& Swinney 1999, \textcolor{black}{Dong 2007}, Brauckmann \& Eckhardt 2017). 
%but the reasons for this have not been elucidated mathematically.
\textcolor{black}{However, this fact is not widely known, probably because the mathematical reasons behind it has not been elucidated.}

\section{High wavenumber Taylor vortices}

The aim of this section is to examine the asymptotic behaviour of the solutions at the first and second peaks seen in figure 3a. To this end, in figure 8, we performed similar calculations at two higher Taylor numbers. The results are summarised using the theoretical scaling to be derived in this section. Essentially, the scaling of $k$ and $Nu$ represent the width of the roll cell in the $z$ direction and the thickness of the boundary layer adjacent to the cylinder wall, respectively. The computations of the solutions are not easy, especially around the first peak, where very thin boundary layers need to be resolved. Although the convergence of the numerical solutions to the asymptotic states is still not perfect, the theoretical scalings are consistent with the overall features of the numerical data.
\begin{figure}
\centering
\begin{tabular}
[c]{cc}
(a)\hspace{10mm}& (b)\hspace{10mm}\\
\includegraphics[scale=1.]{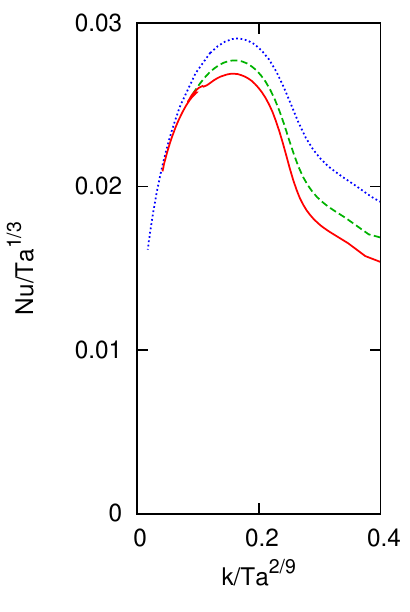} &
\includegraphics[scale=1.]{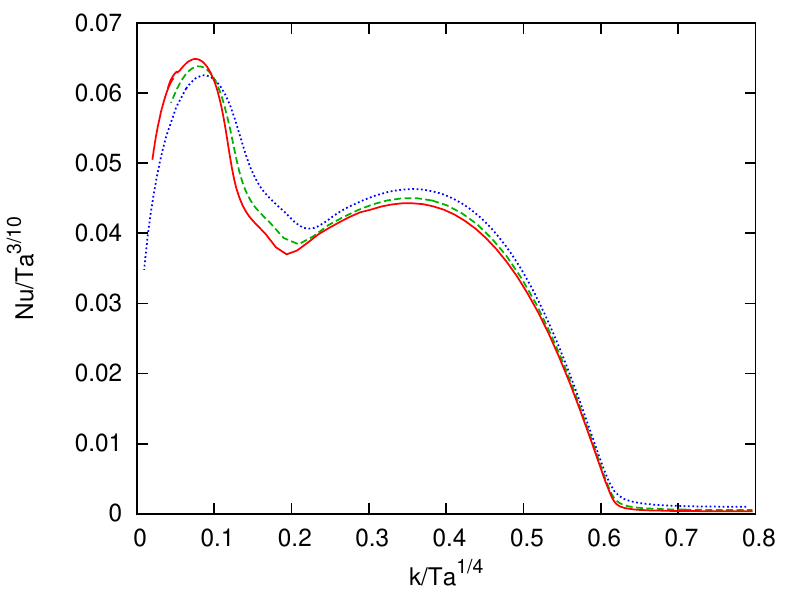}
\end{tabular}
\caption{
Change in the Nusselt number of the Taylor vortex solution when the wavenumber is varied. (a) and (b) use the same numerical results. The outer cylinder is stationary ($a=0$), and the radius ratio is $\eta=5/7$. Red solid, green dashed, and blue dotted curves correspond to $Ta=2.95\times 10^{11}$, $Ta=7.37\times 10^{10}$, and $Ta=9.75\times 10^9$, respectively. 
\textcolor{black}{The blue dotted curve is the same as that shown in figure 3.}
The three crosses in figure 1 are taken from the maxima seen in (a).
%Red solid $R_i=4.4\times 10^5$ ($Ta=2.95\times 10^{11}$), 
%green dashed: $R_i=2.2\times 10^5$ ($Ta=7.37\times 10^{10}$), 
%blue dotted: $R_i=8\times 10^4$ ($Ta=9.75\times 10^9$).
%The flow fields at the left and right peaks are shown in figure 2b and 2c, respectively.
}
\label{fig}
\end{figure}

\begin{figure}
\centering
\begin{tabular}
[c]{cc}
(a)\hspace{10mm}& (b)\hspace{10mm}\\
\includegraphics[scale=0.9]{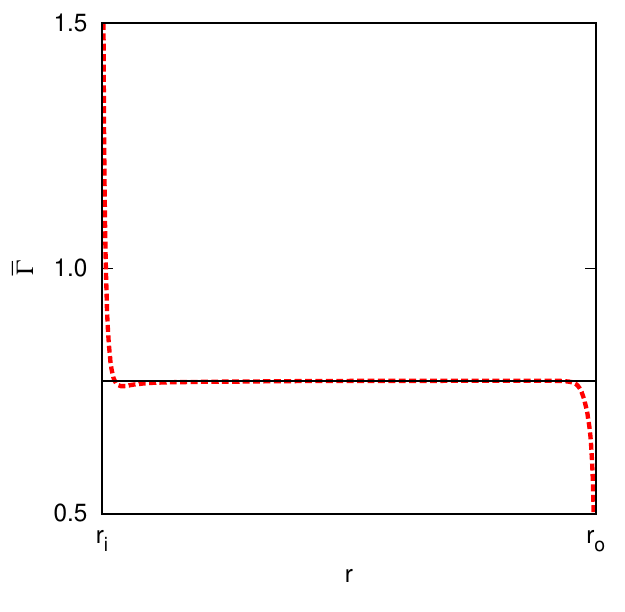} &
\includegraphics[scale=0.9]{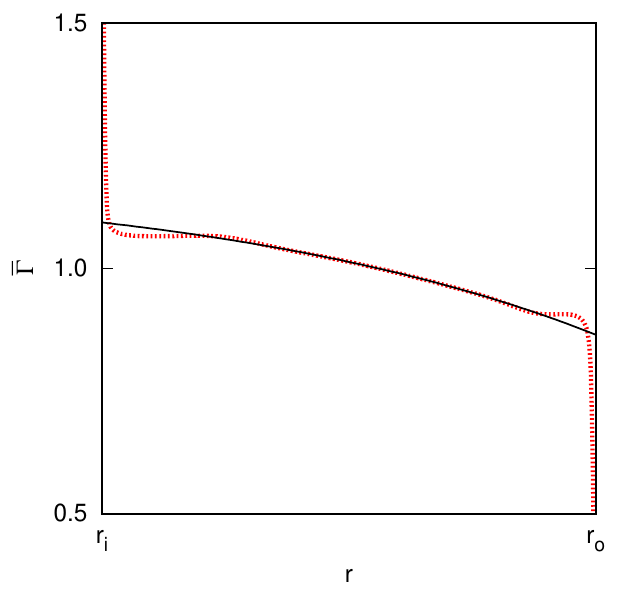}
\end{tabular}
\caption{
Mean flows for the solutions at the extrema of $Nu$ seen in figure 8 ($\eta=5/7, a=0$).
%$R_i=4.4\times 10^5$ ($Ta=2.95\times 10^{11}$). 
The red curves are the numerical results at $Ta=2.95\times 10^{11}$.
(a) The first peak. The black solid line is the asymptotic result $\overline{\Gamma}=\gamma_0$. 
The value of $\gamma_0=0.771$ is estimated at the mid gap. 
%, $\gamma_0=0.771$, 
(b): The second peak. The black curve is the asymptotic result (\ref{transmeanA}).
The value of $A=335.5$ is estimated at the mid gap (see (\ref{Aestimated})).
% $A=335.5$.
}
\label{fig}
\end{figure}

The structure of the flow field in both peaks can be roughly divided into a boundary layer near the wall and a core region in the middle of the gap, as we have seen in figures 2 and 3. On closer inspection, one further notices that the asymptotic structures of the two flows are quite different. For example, in the first peak solution, $\overline{\Gamma}$ has a flat profile in the core region (figure 9a), but this is not the case in the second peak solution (figure 9b). Figure 10 examines how the core flows develop from the near wall region adjacent to the inner cylinder. In the first peak solution, the near-wall structure is somewhat similar to the $k=O(1)$ case, with the wall boundary layer becoming a sharp plume as rounding the corner (top panel). On the other hand, in the second peak solution, no apparent plume can be recognised away from the wall, and the flow only varies slowly in the $z$-direction (bottom panel). The core flow inherits this property, as seen from figure 11, where the axial structure of the fluctuation fields is plotted at the midgap $r=r_m=(r_o+r_i)/2$. For the second peak solution, only a single Fourier mode plays a major role in the core, while for the first peak, a number of harmonics participate in forming the plume. A theoretical explanation of those differences will be deduced below, together with the detailed scalings of the flows.

\begin{figure}
\centering
\includegraphics[scale=0.4]{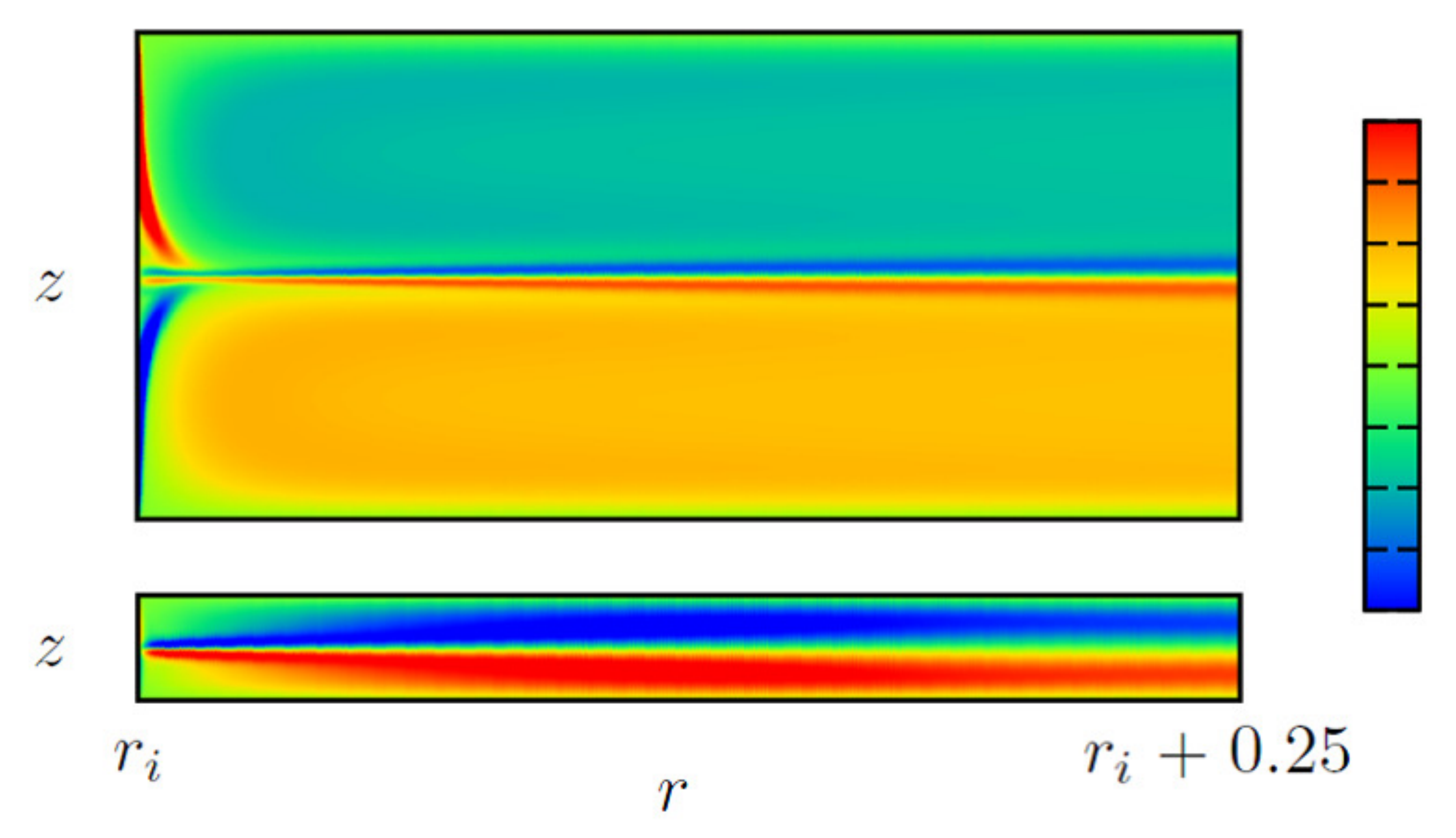}
\caption{The colour map of $\omega/r$
for the first peak (top panel) and the second peak (bottom panel) solutions seen in figure 8. $Ta=2.95\times 10^{11}$. \textcolor{black}{The centre of the colour bar is zero.}
%$\eta=5/7, a=0$.
%The colour bar range is [-1600000,1600000]. 
%$R_i=4.4\times 10^5$ ($Ta=2.95\times 10^{11}$).
}
\label{fig}
\end{figure}

\begin{figure}
\centering
\begin{tabular}
[c]{cc}
(a) first peak \hspace{10mm}& (b) second peak \hspace{10mm}\\
\includegraphics[scale=0.9]{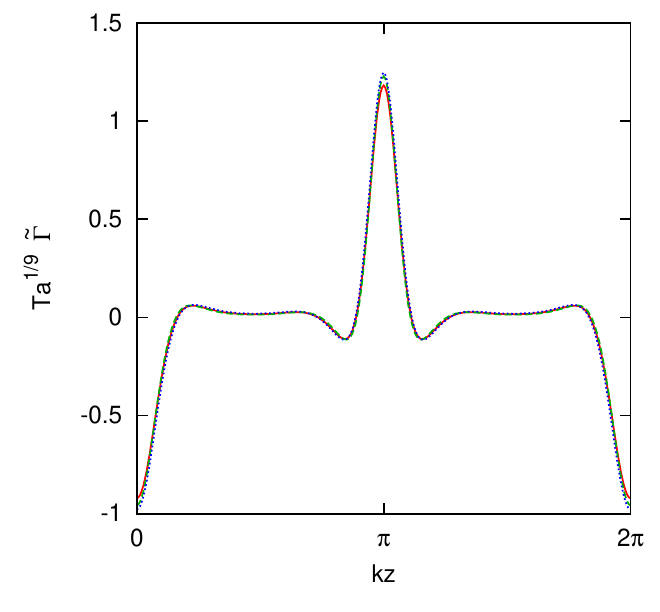} &
\includegraphics[scale=0.9]{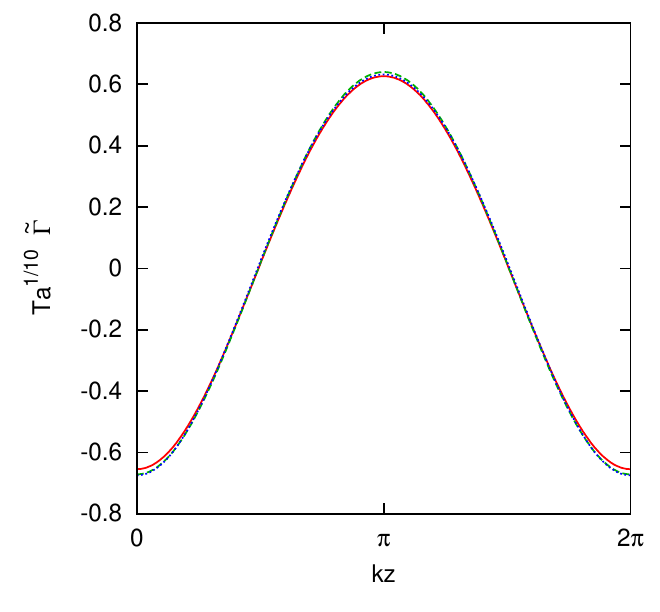}
\end{tabular}
\begin{tabular}
[c]{cc}
(c) first peak \hspace{10mm}& (d) second peak\hspace{10mm}\\
\includegraphics[scale=0.9]{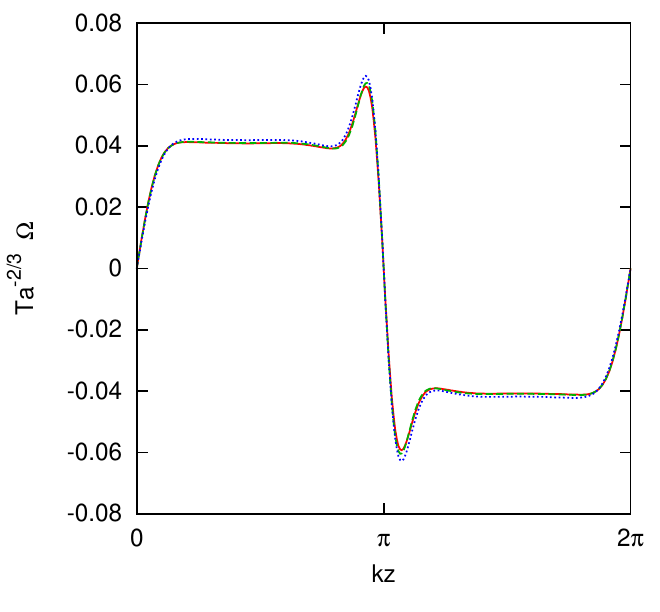} &
\includegraphics[scale=0.9]{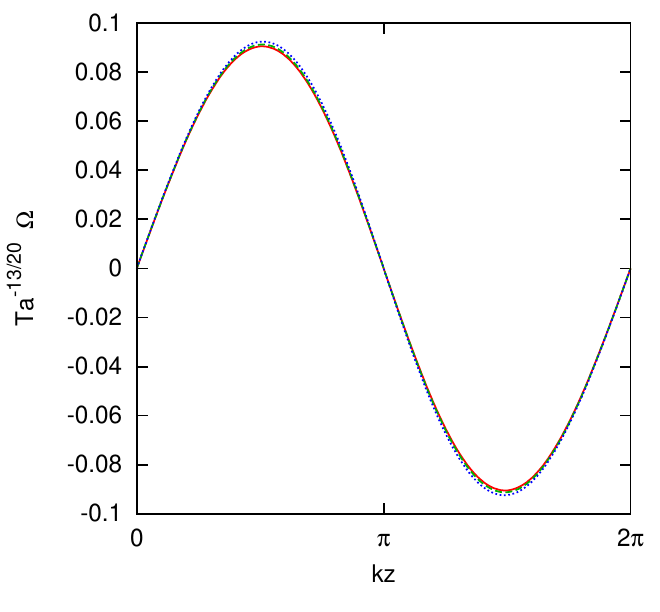}
\end{tabular}
\caption{
The flow fields at the mid gap $r=r_m$ for the first peak and second peak solutions in figure 8.
The red solid, green dashed, and blue dotted curves correspond to $Ta=2.95\times 10^{11}$, $Ta=7.37\times 10^{10}$ and $Ta=9.75\times 10^9$, respectively. 
\textcolor{black}{Note that the value of $k$ is determined by the optimisation of $Nu$ and therefore varies from solution to solution.
The difference in the structure of the first and second peak solutions can be more clearly seen in figure 10, where the axial coordinates are not scaled.}
%Red: $R_i=4.4\times 10^5$ ($Ta=2.95\times 10^{11}$), 
%green: $R_i=2.2\times 10^5$ ($Ta=7.37\times 10^{10}$), 
%blue: $R_i=8\times 10^4$ ($Ta=9.75\times 10^9$).
}
\label{fig}
\end{figure}

\subsection{The first peak asymptotic structure}
Now let us find the scaling of the first peak solution. Hereafter $\delta=1/k$ denotes the length scale of the cell in the $z$ direction. The flow can be divided into core and near-wall zones, as shown in figure 5c. Within the regions $O(\delta)$ away from the cylinder walls, similar asymptotic arguments as the $k=O(1)$ case would apply. Figure 5b shows an enlarged view of the near wall zone where a thin boundary layer, of thickness $\epsilon\ll \delta$ say, emerges around the inviscid region.

The size of the streamfunction in the inviscid region must be the same as that of the core, $\Psi|_c$. Thus the size of the streamfunction within the boundary layer can be found as $O(\Psi|_b)=O(\epsilon \delta^{-1} \Psi|_c)$, noting that the boundary layer thickness relative to the near wall region is $\epsilon/\delta$. Further using the viscous-convective balance in the boundary layer $O(\Psi|_b)=O(\epsilon^{-1}\delta)$, we can find $O(\Psi|_c)=O(\delta^{2}\epsilon^{-2})$. When $\overline{\Gamma}|_c$ is almost constant, the viscous-convective balance in the core is satisfied if $O(\Psi|_c)=O(\delta^{-1})$. Hence, from $O(\Psi|_c)=O(\delta^{2}\epsilon^{-2})$ obtained in the near wall analysis, the two small perturbation parameters are related as $\epsilon=\delta^{3/2}$.

Within the near-wall zone, the plume coming out of the wall boundary layer is so thin that the Coriolis forces and viscous terms cannot balance, unlike the $k=O(1)$ case. This means that the Coriolis force effect should appear as the plume diffuses towards the core region.. The viscous-Coriolis balance in the core can be described as $O(\Psi|_c \delta^{-4})=O(Ta\, \delta^{-1}\widetilde{\Gamma}|_c)$. To estimate $O(\widetilde{\Gamma}|_c)$, we can use the balance $O(\widetilde{\Gamma} |_c\Psi|_c)=O(\Gamma |_b\Psi|_b)$ which will be justified shortly. Using the argument of the inviscid corner $O(\Gamma|_b)=O(1)$. Therefore we have the estimation $O(\widetilde{\Gamma}|_c)=O(\epsilon \delta^{-1})$, which is consistent with the momentum transport balance $O((\partial_z \Psi|_c)\widetilde{\Gamma}|_c)=O(Nu)=O(\epsilon^{-1})$ deduced from (\ref{meanGeq}). This is the final key to unlock the scaling $\delta=O(Ta^{-2/9})$ and $\epsilon=O(Ta^{-1/3})$ with the latter motivated us to use the $Nu\propto Ta^{1/3}$ scaling in figure 8a. If the wind Reynolds number $Re_w$ is defined as the magnitude of the wall-normal velocity component, $Re_w=O(\delta^{-1}\Psi|_c)=O(Ta^{4/9})$.

The asymptotic analysis in the core is summarised as follows. First, we choose $\delta$ as a small perturbation parameter and rescale the axial coordinate as $Z=\delta^{-1} z$. From the scaling argument, the Taylor number has the expansion $Ta=\delta^{-9/2}T_0+\cdots$. Using the core expansions 
\begin{eqnarray}
\Psi=\delta^{-1} \Psi^{(c)}(r,Z)+\cdots,~~
\Gamma=\gamma_0+\delta^{1/2} \Gamma^{(c)}(r,Z)+\cdots
\end{eqnarray}
with a constant $\gamma_0$ to (\ref{Taeq}) yield the leading order equations
\begin{eqnarray}
(\Psi_{r}^{(c)}\partial_Z-\Psi_{Z}^{(c)}\partial_{r})\widetilde{\Gamma}^{(c)}=r_i\widetilde{\Gamma}_{ZZ}^{(c)},\\
(\Psi_{r}^{(c)}\partial_Z-\Psi_{Z}^{(c)}\partial_{r})\Psi_{ZZ}^{(c)}=r_i\Psi_{ZZZZ}^{(c)}-T_0\frac{2\gamma_0}{r_i}\widetilde{\Gamma}_{Z}^{(c)}.
\end{eqnarray}
Here only the fluctuation part of the equations was extracted. The mean part can be directly obtained from (\ref{meanGeq}) as 
\begin{eqnarray}
-\overline{\Psi_{Z}^{(c)}\widetilde{\Gamma}^{(c)}}=N_0,
\end{eqnarray}
using the scaled Nusselt number $N_0=\frac{16\eta r_i^2}{(1+\eta)^4}\epsilon Nu$.

In the plume near the inner wall, we use the expansion
\begin{eqnarray}
\Psi=\epsilon^{-1}\delta \Psi^{(b)}(l,N)+\cdots,~~
\Gamma=\Gamma^{(b)}(l,N)+\cdots
\end{eqnarray}
together with the stretched variables $l=\frac{r-r_i}{\delta}$ and $N=z/\epsilon$.
The leading order part of (4.1a) takes the form
\begin{eqnarray}
\Psi_l^{(b)} \Gamma_N^{(b)}-\Psi_N^{(b)} \Gamma_l^{(b)}= r_i \Gamma_{NN}^{(b)}
\end{eqnarray}
which is essentially identical to (4.10).
If we consider $\Gamma^{(b)}$ is a function of $\psi=\Psi^{(b)}$ and $l$, the equation becomes
\begin{eqnarray}
- \Gamma_l^{(b)}=r_i (\psi_z \Gamma_{\psi}^{(b)})_{\psi}
\end{eqnarray}
where $\Gamma_{\psi}^{(b)}$ should be small when far enough away from the plume.
By integrating this equation across the plume we get
\begin{eqnarray}
\frac{d}{dl}\left (\int^{\infty}_{-\infty} \Gamma^{(b)}  d\psi \right)  = 0.
\end{eqnarray}
The term in the bracket is conserved during the plume diffusion so that the balance $O(\widetilde{\Gamma} |_c\Psi|_c)=O(\Gamma |_b\Psi|_b)$ must be satisfied. The effect of the plume entering the core has to be described by boundary conditions for (5.2)-(5.3) at $r=r_i,r_o$. The conditions could be written using the Dirac delta function as done in Vynniycky \& Masuda (2013).

We do not solve the asymptotic problem, but the theory well explains the beheviour of the Taylor vortex solution. For example, the numerical results shown in figures 11a, c are summarised using the theoretical core scalings $\Omega|_c=O(\delta^{-2}\Psi|_c)=O(Ta^{2/3})$ and $\widetilde{\Gamma}|_c=O(Ta^{-1/9})$. In figure 8a, the thin black line is the asymptotic approximation $\overline{\Gamma}\approx \gamma_0$ with the estimate $\gamma_0=Ta^{-1/2}r_m \overline{v}(r_m)\approx 0.771$. 
%(the estimate $\gamma_0=Ta^{-1/2}r_m \overline{v}(r_m)\approx 0.771$ is used).

From the numerical data, the Nusselt number at the first peak seems to be roughly approximated by $Nu=0.027\,Ta^{1/3}$.
%While the $k=3$ result shown in figure 1 follows $Nu=0.13\,Ta^{1/4}$ at large Taylor numbers.
\textcolor{black}{The asymptotic line intersect with the $k=3$ curve shown in figure 1 at $Ta\approx 6\times 10^7$, which is not a bad approximation of the transition point between the classical and ultimate turbulence regimes.} %NEED REFINEMENT!!!!!!!!!!!!!!!!!!

\subsection{The second peak asymptotic structure}
The asymptotic structure of the second peak solution is more complex because three layers appear in the vicinity of the walls (see figure 12a). We refer to them as the bottom, middle and top boundary layers, in order from the cylinder wall. Understanding their precise scaling requires a delicate matched asymptotic expansion analysis, which we will leave for the next section. Here we shall look only at how the flow scaling is intuitively determined. The argument below contains one assumption that is not strictly fulfilled, but the resultant scaling is nonetheless correct if some minor logarithmic factor effects are omitted.
\begin{figure}
\begin{center}
  \includegraphics[width=0.8\textwidth]{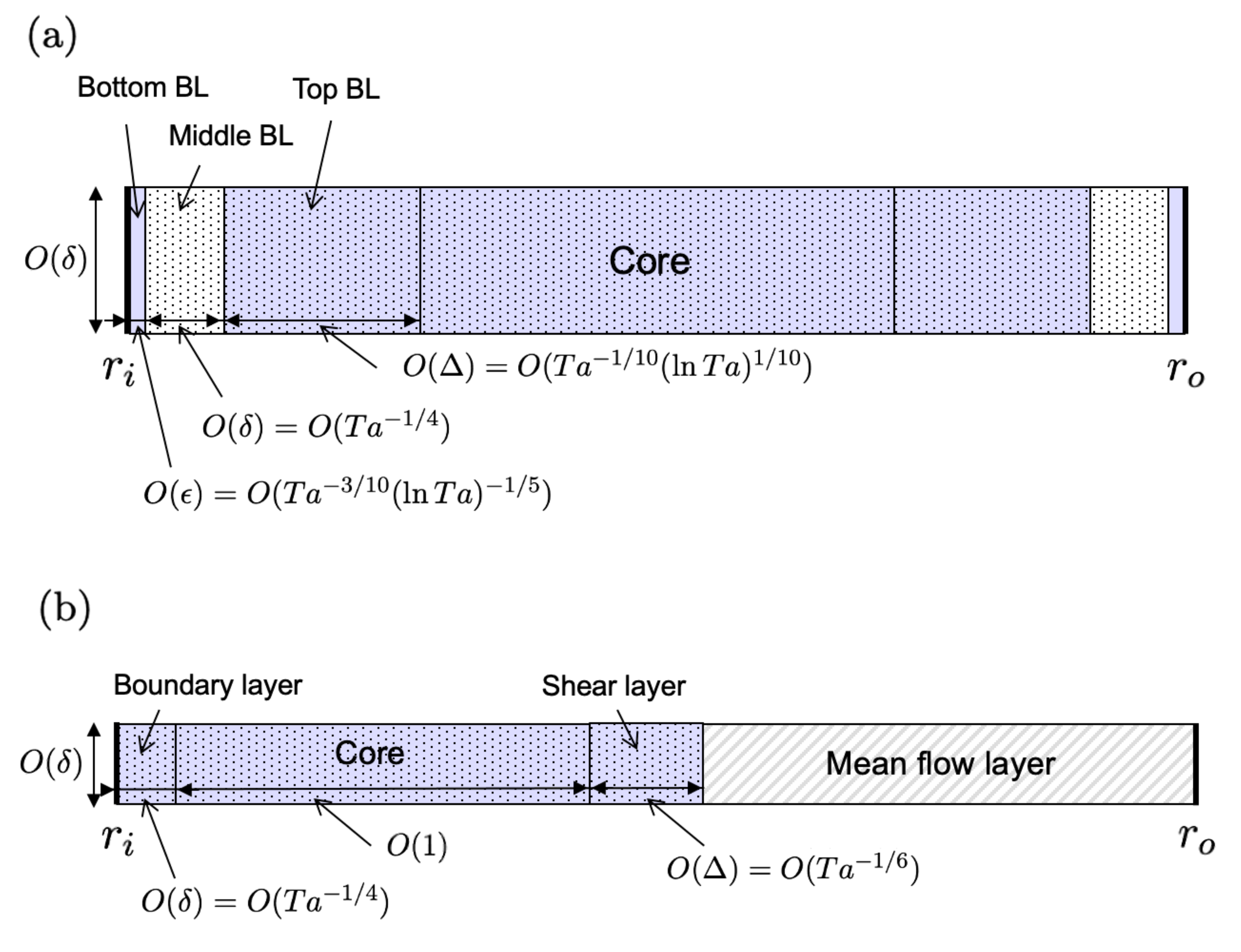}
\end{center}
\caption{
Similar sketch as figure 5 but for the asymptotic states when $k=Ta^{1/4}$. %See figure 5 caption for the 
%In the blue shaded region the viscous effect is not negligible. 
%In the dotted region the Coriolis force is at work. 
(a) The second peak state. BL stands for boundary layer. (b) The transitional state.  The shear layer occurs around the critical radius $r=r_c$.
}
\label{fig}
\end{figure}

The key assumption is $\frac{d\overline{\Gamma}}{dr}=O(1)$ in the core, which did not hold for the first peak case. If the governing equations (\ref{Gammaeq}) and (\ref{Omegaeq}) are linearised around the mean flow, the balances $O(\delta^{-1}\Psi|_c)=O(\delta^{-2}\widetilde{\Gamma})$ and $O(\delta^{-4}\Psi|_c)=O(\delta^{-1}Ta\, \widetilde{\Gamma}|_c)$ must be satisfied. Those balances determine $\delta=O(Ta^{-1/4})$ and $O(\widetilde{\Gamma}|_c)=O(\delta \Psi|_c)$.
Further using $O((\partial_z \Psi|_c)\widetilde{\Gamma}|_c)=O(Nu)=O(\epsilon^{-1})$ deduced from the mean equation (\ref{meanGeq}), the core flow scaling can be written as $O(\Psi|_c)=O(\epsilon^{-1/2})$, $O(\widetilde{\Gamma}|_c)=O(\delta \epsilon^{-1/2})$.

Within the bottom boundary layer, the viscous-convective balance $O(\Psi|_b)=O(\epsilon^{-1} \delta)$ must, of course, be satisfied (the subscript $b$ implies that $\Psi$ is measured at the bottom boundary layer). If we further assume the viscous-Coriolis balance $O(\Psi|_b)=O(Ta\,\epsilon^{4} \delta^{-1})$, we obtain $\epsilon=\delta^{6/5}=Ta^{-3/10}$, which gives $Nu\propto Ta^{3/10}$ in figure 7b. It is this balance that is mostly met, but in fact, is a little broken. As mentioned earlier, the effect of this slight imbalance appears as a logarithmic factor in the scaling of $\epsilon$ in the matched asymptotic expansion. The introduction of the logarithmic factor only has little influence when examining the scaling of numerical data, and so it was ignored in figure 7b. Furthermore, as shown in figures 11b, d, the core scalings $O(\Omega|_c)=O(\delta^{-2}\Psi|_c)=O(Ta^{13/20})$ and $\widetilde{\Gamma}|_c=O(Ta^{-1/10})$ without the logarithmic factor well explain the numerical results. Note that the scaling result suggests $Re_w=O(\delta^{-1}\Psi|_c)=O(\delta^{-1}\epsilon^{-1/2})=O(Ta^{2/5})$.

In the middle boundary layer, the flow coming towards the wall turns back in the region of aspect ratio about unity, so its thickness should be $O(\delta)$. 

The top boundary layer is necessary for the nonlinearity of the fluctuation component to disappear towards the core. Its thickness $\Delta=O(\delta \epsilon^{-1/2})=O(Ta^{-1/10})$ can be obtained by the viscous-convective balance $O(\Delta^{-1}\delta^{-1}\Psi|_t)=O(\delta^{-2})$ and $O(\Psi|_t)=O(\Psi|_c)$, where $O(\Psi|_t)$ is the size of the streamfunction in the top boundary layer. One of the ways to ascertain the presence of the top boundary layer in the visualised flow field is to look at the extreme values of $\Omega$ (see figure 9, bottom panel) or $\widetilde{\Gamma}$. Figure 13a shows the scaled fluctuation $\widetilde{\Gamma}$ at the plume position $z=\pi/k$. It can be seen that the minima approach the wall as the Taylor number increases. In figure 13b, the distance between the wall and the minima is scaled by the theoretical top boundary layer thickness $\Delta$. As expected, all the curves almost overlap around the minima.

\section{Matched asymptotic expansion when $k=O(Ta^{1/4})$}

\begin{figure}
\centering
\begin{tabular}
[c]{cc}
(a)\hspace{10mm}& (b)\hspace{10mm}\\
\includegraphics[scale=0.9]{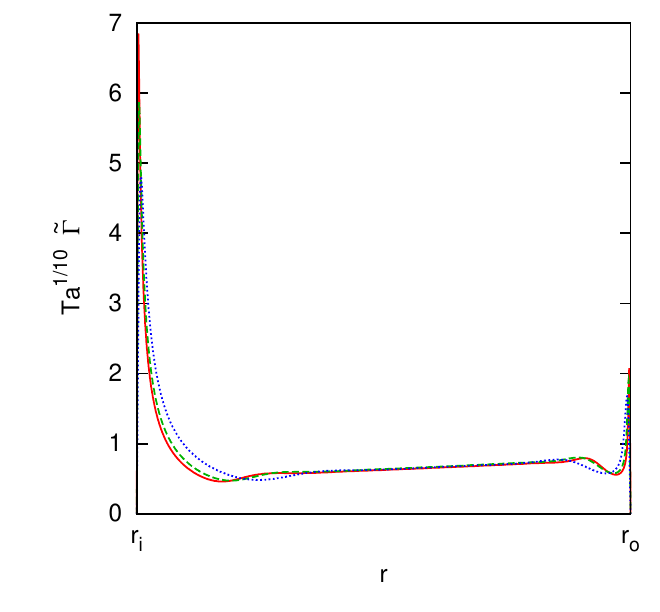} &
\includegraphics[scale=0.9]{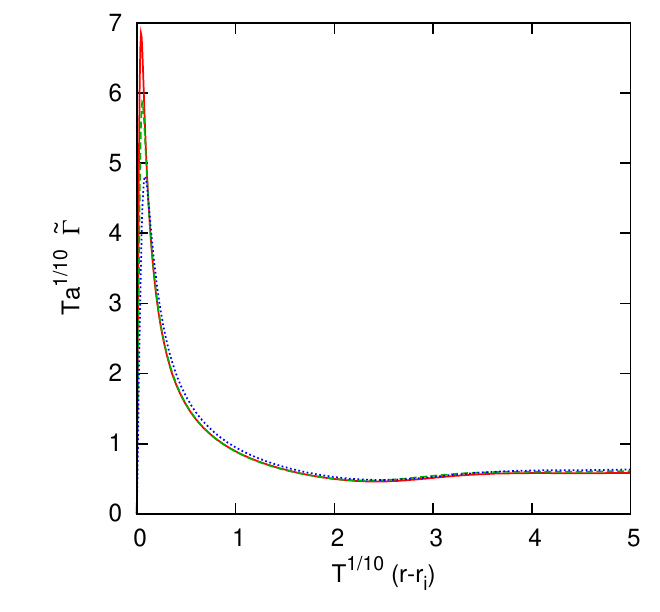}
\end{tabular}
\caption{
The profile of $\widetilde{\Gamma}$ at the plume position $z=\pi/k$ for the Taylor vortex solution at the second peak.
 $\eta=5/7, a=0$.
The red solid, green dashed, and blue dotted curves correspond to $Ta=2.95\times 10^{11}$, $Ta=7.37\times 10^{10}$ and $Ta=9.75\times 10^9$, respectively.
}
\label{fig}
\end{figure}

The goal of this section is to derive a matched asymptotic expansion consistent with the behaviour of the second peak solutions. This asymptotic state can be reached by decreasing the wavenumber from the linear critical point, as seen in figure 8b. However, for $k/Ta^{1/4}$ ranging from about 0.6 to 0.8, the appropriate scaling of the Nusselt number is $O(Ta^0)$, which is different from that observed for the second peak state; see figure 14a. We shall study this transitional state in section 6.1, followed by the analysis of the second peak in section 6.2.

The most important previous work throughout this section is due to Denier (1992), who applied the Hall \& Lakin (1988) type high wavenumber asymptotic theory to Taylor-Couette flow in the narrow gap limit. As briefly commented in that paper, the extension of the theory to wide-gap cases should be straightforward. In section 6.1, we derive an analytic expression of the Nusselt number and compare it with the numerical solution for the first time. 

As the solution moves away from the linear critical point in the transitional state, the vortex grows from the vicinity of the inner cylinder (figure 14c). The asymptotic state corresponding to the second peak appears when the vortex fills the entire gap. A similar scenario was suggested in Denier (1992) using a matched asymptotic expansion, but the scaling obtained is unfortunately not consistent with the behaviour of the numerical solutions. The main reason for this is that the matching is not actually possible in the two-layer boundary layer structure assumed in Denier (1992). In section 6.2, we will resolve that difficulty by modifying the structure of the near wall-zone to three layers.

We choose $\delta=k^{-1}$ as a small perturbation parameter and introduce the scaled axial variable $Z=\delta^{-1}z$. According to Hall (1982), the linear critical point of curved flow problems typically has the Taylor number expansion $Ta=\delta^{-4} T_0+\cdots$. This is also true for Taylor Couette flow from the linear critical point to the second peak.

\begin{figure}
\centering
\begin{tabular}
[c]{cc}
(a)\hspace{10mm}& (b)\hspace{10mm}\\
\includegraphics[scale=0.9]{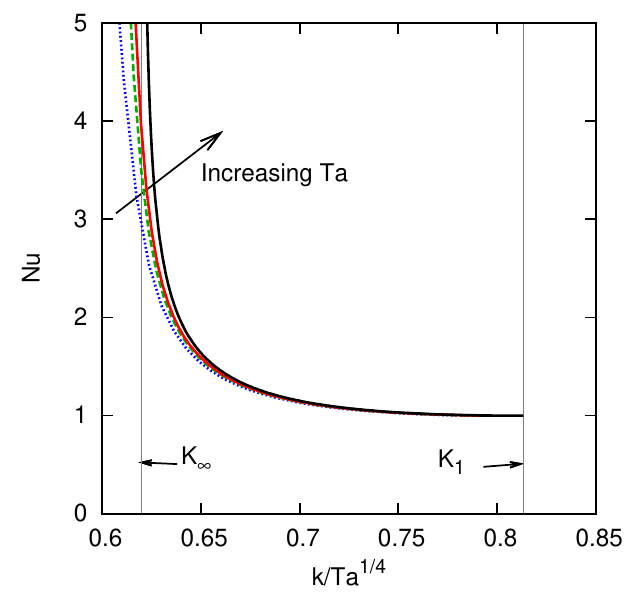} &
\includegraphics[scale=0.9]{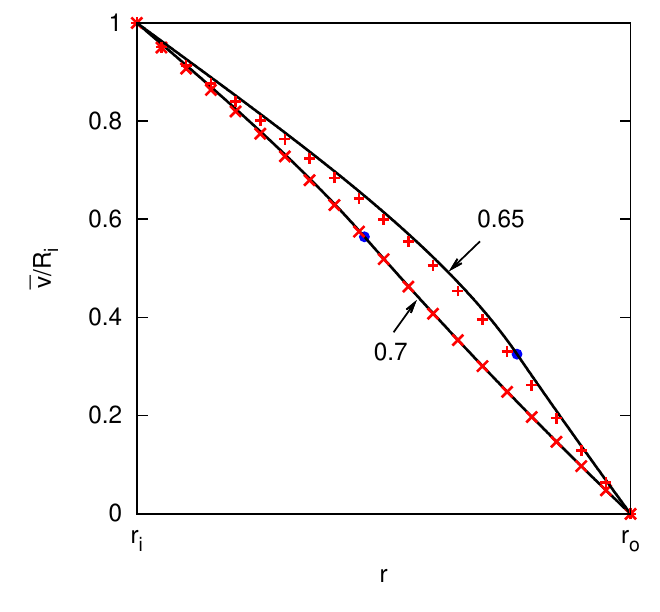}
\end{tabular}
\\
(c)~~~~~~~~~~~~~~~~~~~~~~~~~~~~~~~~~~~~~~~~~~~~~~~~~~~~~~~~~~~~~~~~~~~~~~~~~~~~~~~~~~~\\
\includegraphics[scale=0.4]{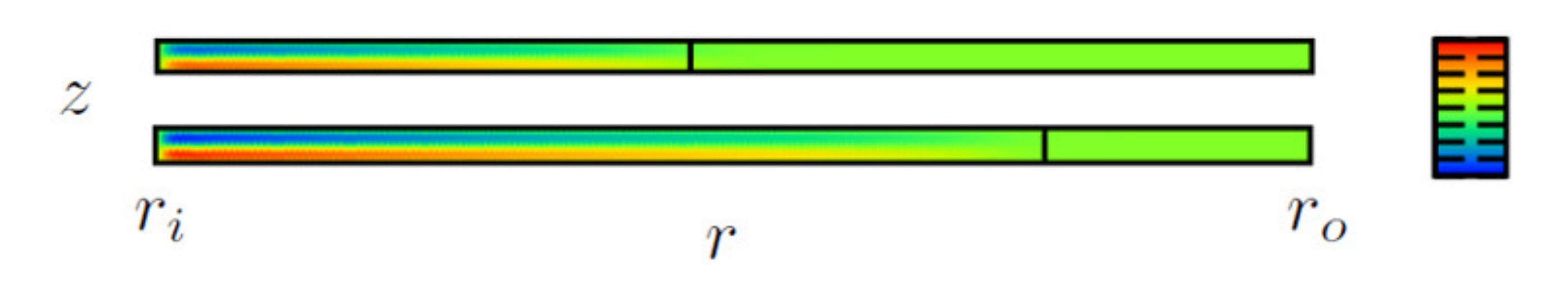}
\caption{(a): The close-up of figure 8b around the bifurcation point. 
The black solid curve is the asymptotic result (\ref{transNuNu}), which has the property that $Nu=1$ at $k/Ta^{1/4}=K_1$, and that $Nu\rightarrow \infty$ as $k/Ta^{1/4}\rightarrow K_{\infty}$.
(b) The mean flow profile at $k/Ta^{1/4}=0.65, 0.7$. The black solid curves are the asymptotic result (\ref{meanflowfinal}). 
The bullets indicate $r=r_c$ given in (\ref{transrc}). 
The red points are the numerical Taylor vortex result for $Ta=9.75\times 10^9$.
%$R_i=8\times 10^4$. 
 (c) The colourmap of $\omega/r$ for the numerical Taylor vortex.
The top and bottom panels correspond to $k/Ta^{1/4}=0.7$ and 0.65, respectively.
Tics are placed at $r=r_c$.
%Top (bottom) panel is the solution with $k/Ta^{0.25}=0.65$ (0.7), where the colour bar range is [-320000,320000] ([-400000,400000]). 
}
\label{fig}
\end{figure}
\subsection{The transitional states around the linear critical point}
For simplicity, we fix the outer cylinder (i.e. $a=\Gamma_o=0$) as Denier (1992). The analysis of the general cases ($a\neq 0$) \textcolor{black}{is relegated to} Appendix B as it is somewhat complex. The asymptotic structure of the transitional state is depicted in figure 12b. The vortices are concentrated in the core region $r_i<r<r_c$, where $r_c \in [r_i,r_o]$ is the critical radius to be determined analytically. We can show that the vortex amplitude decays exponentially within a shear layer of thickness $O(\Delta)=O(Ta^{-1/6})$ appearing around the critical radius. The structure of the shear layer is identical to that described in Denier (1992) and is not discussed in detail here. The only important fact that will be used later is that the mean flow and its derivative are continuous across this layer.

In the region where $r$ is greater than $r_c$ it is sufficient to analyse the mean flow. To leading order, the left-hand side of (\ref{meanGeq}) can be ignored and hence the mean flow is a linear combination of a constant and $r^2$.
Writing 
\begin{eqnarray}
N_0=\frac{16\eta r_i^2}{(1+\eta)^4}Nu\label{transN0}
\end{eqnarray}
 for simplicity, we get the asymptotic approximation
\begin{eqnarray}
\overline{\Gamma}=B(r_o^2-r^2)+\cdots,\qquad B=\frac{N_0}{2r_o^2}.\label{transmeanB}
\end{eqnarray}
Note that we used the fact that $\overline{\Gamma}$ should vanish at $r=r_o$.

In the core region, the appropriate expansions are
\begin{eqnarray}
\Psi=\widehat{\Psi}^{(c)}(r)\sin Z+\cdots,~~
\widetilde{\Gamma}=\delta \widehat{\Gamma}^{(c)}(r)\cos Z+\cdots,~~
\overline{\Gamma}=\overline{\Gamma}^{(c)}(r)+\cdots.
\end{eqnarray}
Substituting them into (\ref{Taeq}), the leading order parts yield
\begin{eqnarray}
\overline{\Gamma}_r^{(c)} \widehat{\Psi}^{(c)} =r\widehat{\Gamma}^{(c)}, \qquad 0=\widehat{\Psi}^{(c)}+T_0 \frac{2\overline{\Gamma}^{(c)}}{r^2}\widehat{\Gamma}^{(c)}.\label{transcore}
\end{eqnarray}
For those equations to have a nontrivial solution $0=r^3+2T_0\overline{\Gamma}^{(c)}\overline{\Gamma}^{(c)}_r$ must be satisfied. Therefore the mean flow is obtained as
\begin{eqnarray}
\overline{\Gamma}^{(c)}(r)=\sqrt{\frac{A-r^4}{4T_0}}.\label{transmeanA}
\end{eqnarray}
The constant $A$ must be
\begin{eqnarray}
A=r_i^4+4T_0\Gamma_i^2\label{transA}
\end{eqnarray}
to satisfy $\overline{\Gamma}^{(c)}(r_i)=\Gamma_i$.

A boundary layer of thickness $O(\delta)$ is needed near the inner cylinder to satisfy the no-slip boundary conditions. 
In this layer we introduce the stretched variable $\xi=\frac{r-r_i}{\delta}$ and assume the expansion
\begin{eqnarray}
\Psi=\Psi^{(b)}(\xi,Z)+\cdots,~~
\Gamma= \Gamma_i+\delta\Gamma^{(b)}(\xi,Z)+\cdots.
\end{eqnarray}
The leading order equations are obtained as
\begin{eqnarray}
(\Psi_{\xi}^{(b)}\partial_Z-\Psi_{Z}^{(b)}\partial_{\xi})\Gamma^{(b)}=r_i(\partial_{\xi}^2+\partial_Z^2)\Gamma^{(b)},\\
(\Psi_{\xi}^{(b)}\partial_Z-\Psi_{Z}^{(b)}\partial_{\xi})(\partial_{\xi}^2+\partial_Z^2)\Psi^{(b)}=r_i(\partial_{\xi}^2+\partial_Z^2)^2\Psi^{(b)}-T_0\frac{2\Gamma_i\Gamma_{Z}^{(b)}}{r_i}.
\end{eqnarray}
The no-slip boundary conditions are 
\begin{eqnarray}
\Psi^{(b)}=\Psi_{\xi}^{(b)}=0,\qquad \Gamma^{(b)}=0\qquad \text{at} \qquad \xi=0,
\end{eqnarray}
while for the far-field $\xi\rightarrow \infty$ we use the matching conditions
\begin{eqnarray}
\Psi^{(b)}\rightarrow \widehat{\Psi}^{(c)}(r_i)\sin Z,\qquad \widetilde{\Gamma}^{(b)}  \rightarrow \widehat{\Gamma}^{(c)}(r_i)\cos Z,\qquad \overline{\Gamma}^{(b)}\rightarrow \overline{\Gamma}^{(c)}_r(r_i)\xi.
\end{eqnarray}

Finally, we determine the unknown constants $B$ and $r_c$ from the continuity
of $\overline{\Gamma}$ and $r^3(r^{-2}\overline{\Gamma})_r$ at $r=r_c$.
Using the mean flows (\ref{transmeanA}) and (\ref{transmeanB}), the continuity conditions can be written as 
\begin{eqnarray}
\sqrt{\frac{A-r_c^4}{4T_0}}=B(r_o^2-r_c^2),\qquad \frac{A}{\sqrt{T_0(A-r_c^4)}}=2Br_o^2,
\end{eqnarray}
where $A$ is known by (\ref{transA}). 
From those equations it is easy to find
\begin{eqnarray}
r_c=\sqrt{\frac{r_i^4+4T_0\Gamma_i^2}{r_o^2}},\label{transrc} \\
\frac{1}{B}=\sqrt{4T_0\left (\frac{r_o^4}{r_i^4+4T_0\Gamma_i^2}-1 \right )}.\label{transBinv}
\end{eqnarray}
Thus if $T_0$ is given the leading order mean flow is completely determined as
\begin{eqnarray}
\overline{\Gamma}\approx 
\left \{
\begin{array}{c}
(r_o^2-r^2)\sqrt{\frac{r_i^4+4T_0\Gamma_i^2}{4T_0\left (r_o^4-r_i^4-4T_0\Gamma_i^2 \right )}}\qquad \text{if} \qquad r> r_c,\\
\sqrt{\frac{r_i^4-r^4+4T_0\Gamma_i^2}{4T_0}}\qquad \text{if} \qquad r\leq r_c.
\end{array}
\right . \label{meanflowfinal}
\end{eqnarray}
This is the black curve in figure 14b, which predicts the numerical results very well.
Note that the analytic expression of $\widehat{\Psi}^{(c)}$ and $\widehat{\Gamma}^{(c)}$ can also be found by (\ref{transcore}) and the leading order part of the mean flow equation (\ref{meanGeq})
\begin{eqnarray}
-\frac{\overline{\Gamma}^{(c)}_r}{2r}(\widehat{\Psi}^{(c)})^2=r^3(r^{-2}\overline{\Gamma}^{(c)})_r+N_0. \label{Demean}
\end{eqnarray}

The Nusselt number is found by (\ref{transN0}), (\ref{transmeanB}), (\ref{transBinv}) as
\begin{eqnarray}
Nu(T_0)=\frac{(1+\eta)^4}{16\eta^3\sqrt{T_0}}\sqrt{\frac{r_i^4+4T_0\Gamma_i^2}{r_o^4-r_i^4-4T_0\Gamma_i^2}}.\label{transNuNu}
\end{eqnarray}
This function is plotted by the black curve in figure 14a using $k/Ta^{1/4}= T_0^{-1/4}$ for the horizontal axis. The other curves in the figure, corresponding to the Taylor vortex solutions at finite $Ta$, clearly approach the asymptotic result as $Ta\rightarrow \infty$.

As the scaled wavenumber decreases from the linear critical point $K_1=T_1^{-1/4}$, the Nusselt number increases from the laminar value unity. 
The asymptotic result $T_1=\frac{r_i^2(r_o^2-r_i^2)}{4\Gamma_i^2}$ can be found from (\ref{transrc}) by setting $r_c=r_i$ and $T_0=T_1$.

Similar to the narrow gap case studied by Denier et al. (1992), $Nu$ blows up at finite $T_0$.
Writing $K_{\infty}=T_{\infty}^{-1/4}$, we can deduce $T_{\infty}=\frac{r_o^4-r_i^4}{4\Gamma_i^2}$ from (\ref{transrc}) by using $r_c=r_o$ and $T_0=T_{\infty}$. 
That is, in figure 14a, $K_{\infty}$ is the point at which the flow switches from the transitional state to the second peak state.

\subsection{The second peak states}
We now turn our attention to the second peak asymptotic states. 
Again $\delta$ is the small asymptotic parameter, and $Ta=\delta^{-4}T_0+\cdots$. For the bottom boundary layer thickness $\epsilon$, we impose the condition
\begin{eqnarray}
\frac{\delta^6}{\epsilon^{5}}=\ln\frac{\delta}{\epsilon},\label{epsdelta}
\end{eqnarray}
which is needed for successful matching.
The boundary layer thickness is estimated as $O(\epsilon)=O(Ta^{-3/10}(\ln Ta)^{-1/5})$. Therefore the Nusselt number and the wind Reynolds number scalings now involve the logarithmic correction as $Nu=O(\epsilon^{-1})=O(Ta^{3/10}(\ln Ta)^{1/5})$, $Re_w=O(\delta^{-1}\Psi|_c)=O(\delta^{-1}\epsilon^{-1/2})=O(Ta^{2/5}(\ln Ta)^{1/10})$.

Let us start with the core analysis by writing
\begin{eqnarray}
\Psi=\epsilon^{-1/2} \widehat{\Psi}^{(c)}(r)\sin Z+\cdots,~~\\
\widetilde{\Gamma}=\delta \epsilon^{-1/2} \widehat{\Gamma}^{(c)}(r)\cos Z+\cdots,~~
\overline{\Gamma}=\overline{\Gamma}^{(c)}(r)+\cdots.
\end{eqnarray}
Those expansions, based on the argument in section 5, yield the identical leading order equations (\ref{transcore}) as the transitional state. 
As a result, the mean flow is given by (\ref{transmeanA}). However, for the second peak state, the multi-layered near-wall structure must also be present near the outer cylinder, and thus there is no way to determine the constant $A$ a priori. 
The leading order part of the mean flow equation (\ref{meanGeq}) is obtained as
\begin{eqnarray}
-\frac{\overline{\Gamma}^{(c)}_r}{2r}(\widehat{\Psi}^{(c)})^2=N_0, \label{2ndmean}
\end{eqnarray}
where $N_0=\frac{16\eta r_i^2}{(1+\eta)^4}\epsilon Nu$.
The unknowns appeared in the core solutions $\overline{\Gamma}^{(c)}, \widehat{\Psi}^{(c)}$, and $\widehat{\Gamma}^{(c)}$ are $A$ and $N_0$.

The two near-wall regions have identical asymptotic structure, so only the layers near the inner cylinder are examined below.
The top boundary layer expansions are
\begin{eqnarray}
\Psi=\epsilon^{-1/2} \Psi^{(t)}(\zeta,Z)+\cdots,~~
\Gamma=\gamma_i+\delta \epsilon^{-1/2} \Gamma^{(t)}(\zeta,Z)+\cdots,
\end{eqnarray}
where the constant $\gamma_i=\overline{\Gamma}^{(c)}(r_i)$ comes from the leading order core mean flow.
The stretched variable $\zeta=\frac{r-r_i}{\Delta}$ is defined by using the top boundary layer thickness $\Delta=\delta\epsilon^{-1/2}=O(Ta^{-1/10}(\ln Ta)^{1/10})$. 
The leading order governing equations for the top boundary layer are similar to those for the core region in the first peak states:
\begin{eqnarray}
(\Psi_{\zeta}^{(t)}\partial_Z-\Psi_{Z}^{(t)}\partial_{\zeta})\widetilde{\Gamma}_t=r_i\widetilde{\Gamma}_{ZZ}^{(t)},\\
(\Psi_{\zeta}^{(t)}\partial_Z-\Psi_{Z}^{(t)}\partial_{\zeta})\Psi_{ZZ}^{(t)}=r_i\Psi_{ZZZZ}^{(t)}-T_0\frac{2\gamma_i}{r_i}\widetilde{\Gamma}_{Z}^{(t)},\\
-\overline{\Psi_{Z}^{(t)}\widetilde{\Gamma}^{(t)}}=N_0.
\end{eqnarray}
As $\zeta \rightarrow \infty$, the flow matches to the core solution as
\begin{eqnarray}
\Psi^{(t)}\rightarrow \widehat{\Psi}^{(c)}(r_i)\sin Z,\qquad \widetilde{\Gamma}^{(t)} \rightarrow \widehat{\Gamma}^{(c)}(r_i)\cos Z,\qquad \overline{\Gamma}^{(t)}\rightarrow \overline{\Gamma}^{(c)}_r(r_i)\zeta.
\end{eqnarray}
While the asymptotic behaviour of the flow towards the wall should be
\begin{eqnarray}
\Psi^{(t)}\rightarrow \zeta^{1/3}\widehat{\Psi}^{(t)}(Z),\qquad \Gamma^{(t)} \rightarrow \zeta^{-1/3}\widehat{\Gamma}^{(t)}(Z)\qquad \text{as}\qquad \zeta \rightarrow 0,
\end{eqnarray}
which is similar to that used in Denier (1992). However, to match this with the bottom boundary layer, we must insert the middle boundary layer.

Recall that the middle boundary layer is the special place where the radial and axial derivatives of the flow have the same size. 
Thus the expansions there must be written in terms of $\xi=\frac{r-r_i}{\delta}$ as
\begin{eqnarray}
\Psi=\epsilon^{-1/3} \Psi^{(m)}(\xi,Z)+\cdots,~~
\Gamma=\gamma_i+\delta \epsilon^{-2/3} \Gamma^{(m)}(\xi,Z)+\cdots.
\end{eqnarray}
Given the expansions it is a straightforward task to find that the leading order equations 
\begin{eqnarray}
(\Psi_{\xi}^{(m)}\partial_Z-\Psi_{Z}^{(m)}\partial_{\xi})\widetilde{\Gamma}^{(m)}=0,\\
(\Psi_{\xi}^{(m)}\partial_Z-\Psi_{Z}^{(m)}\partial_{\xi})(\partial_{\xi}^2+\partial_Z^2)\Psi^{(m)}=-T_0\frac{2\gamma_i}{r_i}\widetilde{\Gamma}_{Z}^{(m)},\\
-\overline{\Psi_{Z}^{(m)}\widetilde{\Gamma}^{(m)}}=N_0,\label{N0middle}
\end{eqnarray}
are inviscid. Here the Coriolis force term participating in the second equation is critical for successful matching to the top boundary layer:
\begin{eqnarray}
\Psi^{(m)}\rightarrow \xi^{1/3}\widehat{\Psi}^{(t)}(Z),\qquad \Gamma^{(m)} \rightarrow \xi^{-1/3}\widehat{\Gamma}^{(t)}(Z)\qquad \text{as}\qquad \xi\rightarrow \infty.
\end{eqnarray}
On the other hand, the appropriate behaviour of the solution towards the wall $\xi\rightarrow 0$ can be found as 
\begin{eqnarray}
\Psi^{(m)}\rightarrow \xi (-\ln \xi)^{1/3} \widehat{\Psi}^{(b)}(Z),\qquad \Gamma^{(m)} \rightarrow \xi^{-1}(-\ln \xi)^{-1/3}\widehat{\Gamma}^{(b)}(Z)
\end{eqnarray}
by seeking the consistent limiting behaviour of the solution. The key observation here is that for both matching conditions, the rate of increase in $\Psi$ and the rate of decrease in $\Gamma$ are the same when moving away from the wall. This is a requirement to satisfy the angular momentum transport (\ref{N0middle}).

The leading order part of the bottom boundary layer expansions
\begin{eqnarray}
\Psi=\frac{\delta}{\epsilon} \Psi^{(b)}(Y,Z)+\cdots,~~
\Gamma= \Gamma^{(b)}(Y,Z)+\cdots
\end{eqnarray}
matches to the middle boundary layer if (\ref{epsdelta}) holds and 
\begin{eqnarray}
\Psi^{(b)}\rightarrow Y\widehat{\Psi}^{(b)}(Z),\qquad \Gamma^{(b)} \rightarrow Y^{-1}\widehat{\Gamma}^{(b)}(Z)
\end{eqnarray}
as $Y=\frac{r-r_i}{\epsilon} \rightarrow \infty$.
The leading order equations in the bottom boundary layer are
\begin{eqnarray}
(\Psi_{Y}^{(b)}\partial_Z-\Psi_{Z}^{(b)}\partial_{Y})\Gamma^{(b)}=r_i\Gamma_{YY}^{(b)},\\
(\Psi_{Y}^{(b)}\partial_Z-\Psi_{Z}^{(b)}\partial_{Y})\Psi_{YY}^{(b)}=r_i\Psi_{YYYY}^{(b)}.
\end{eqnarray}
Thanks to the radial diffusivity the no-slip boundary conditions
\begin{eqnarray}
\Psi^{(b)}=\Psi_{Y}^{(b)}=0,\qquad \Gamma^{(b)}=\Gamma_i \qquad \text{at} \qquad Y=0
\end{eqnarray}
can be imposed.

Although the combined boundary layers appear rather complex, their role is merely to define the relationship between $A$ and $N_0$. Once $T_0$ and $A$ are fixed, the flow near the inner cylinder can be \textcolor{black}{calculated} to find $N_0=f_i(A,T_0)$. A similar calculation for the near outer cylinder region yields another functional relationship $N_0=f_o(A,T_0)$. Hence, in principle, $A(T_0)$ and $N_0(T_0)$ can be determined by solving the two near-wall regions numerically. We do not perform such challenging calculations since we have already seen in section 5 that the finite $Ta$ numerical results are consistent with the asymptotic theory.

The value of $A$ can be predicted from the finite $Ta$ numerical solutions as
\begin{eqnarray}
A=(2k^{-2}r_m\overline{v}(r_m))^2+r_m^4\label{Aestimated}
\end{eqnarray}
by using the mean flow at the mid gap. The black curve in figure 9b is the result of using this $A$ in (\ref{transmeanA}), and it can be seen that this asymptotic prediction matches the numerical calculation in the entire core region. Similarly, the fluctuation component in the core can be calculated explicitly. The black curve in figure 15 is the asymptotic approximation
\begin{eqnarray}
\widetilde{\Gamma}\approx Ta^{-1/4}r\left  (\frac{32\eta r_i^2}{(1+\eta)^4}\frac{Nu}{\sqrt{(A-r^4)}} \right )^{1/2} \cos Z,\label{asymfluccore}
\end{eqnarray}
which is in good agreement with the numerical result.

\begin{figure}
\begin{center}
  \includegraphics[width=0.8\textwidth]{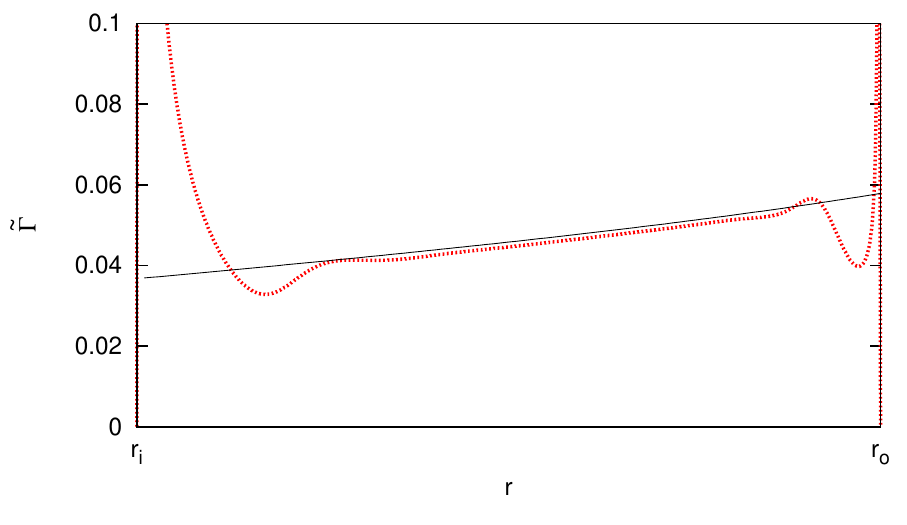}
\end{center}
\caption{The red dotted curve is the profile of $\widetilde{\Gamma}$ at the plume position $z=\pi/k$ for the Taylor vortex solution at the second peak.
The same data as figure 13 ($\eta=5/7,a=0,Ta=2.95\times 10^{11}$). The black curve is the asymptotic result (\ref{asymfluccore}).}
\label{fig}
\end{figure}

\begin{figure}
\centering
\begin{tabular}
[c]{cc}
(a)\hspace{10mm}& (b)\hspace{10mm}\\
\includegraphics[scale=0.9]{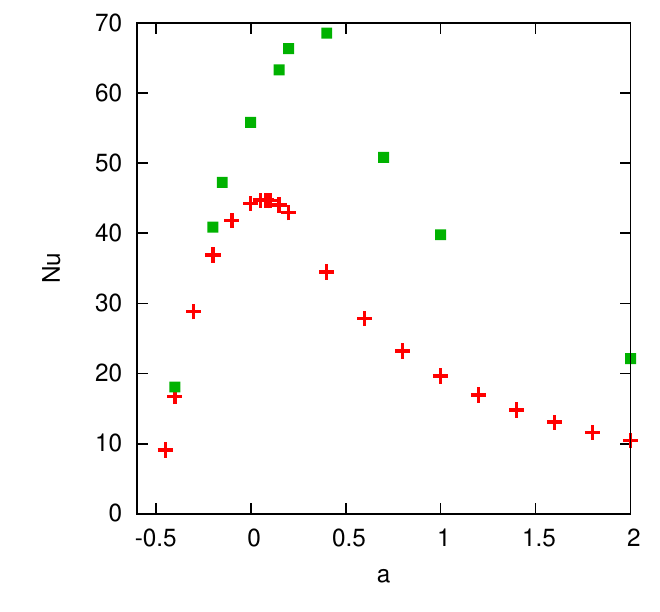} &
\includegraphics[scale=0.9]{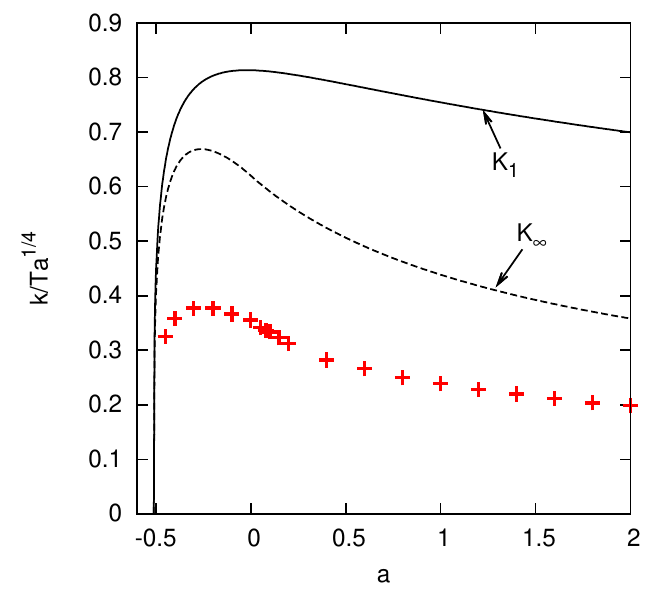}
\end{tabular}
\caption{
 %2nd peak. Rayleigh line $a=-\eta^2\approx -0.51$. 
 (a) The Nusselt number at $Ta=10^{10}$ for $\eta=5/7$.
 The red crosses are the Taylor vortex solution with the optimised wavenumbers shown in (b).
The cyan squares are experimental results by Dennis et al. (2011).
(b) The red crosses are the local maximum of $Nu$ closest to the bifurcation point. 
According to the asymptotic theory, the Taylor vortex solution bifurcates from circular Couette flow at $K_1$ (see (\ref{ApK1})). 
The Nusselt number $Nu$ is $O(1)$ when $k/Ta^{1/4}\in (K_\infty, K_1)$, where $K_{\infty}$ is given in (\ref{ApKinf}). 
For $k/Ta^{1/4}<K_\infty$, the $Nu$ scaling is like the second peak state.
}
\label{fig}
\end{figure}

We shall also comment briefly on the case where $a$ is non-zero. For any $a$, it is not so difficult to decrease $k$ of the Taylor vortex solution branch from the linear critical point until the Nusselt number takes a peak. The red crosses in figure 16a summarise the $Nu$ at the peak for various $a$. The numerical result shows that the optimised $Nu$ is maximised \textcolor{black}{when the outer cylinder slightly counter-rotates} ($a\approx 0.08$). This is qualitatively in line with the experimental ultimate turbulence results shown by cyan squares. However, for the experimental results, the maximum of $Nu$ is attained at $a\approx 0.3$ and, as noted in section 3, the Nusselt number scaling is $Nu\propto Ta^{0.38}$. The Taylor vortex results naturally have a much smaller $Nu$, because it is scaled like $Ta^{0.3}$ as the second peak seen in the $a=0$ case. In figure 16b, we also plotted $K_1$ and $K_{\infty}$ obtained in the transitional state analysis (see (\ref{ApK1}) and (\ref{ApKinf}) in Appendix B). Roughly speaking, the peaks occur when $k/Ta^{1/4}$ is about half of $K_{\infty}$. The linear and nonlinear instabilities disappear at the Rayleigh line $a=-\eta^2\approx -0.51$, which can be found from Rayleigh's stability condition.

Brauckmann et al. (2016) and Brauckmann \& Eckhardt (2017) reported that when the gap is narrow, there are two values of $R_{\Omega}=(1-\eta)\frac{R_i+R_o}{R_i-\eta R_o}$ at which the local maximum of $Nu$ occurs. Whether similar results can be obtained with the Taylor vortex solution is left for future work.

\section{Conclusions and discussion}

In this study, the Taylor vortex solution branch is continued to very high Taylor numbers, and a theoretical rationale is given for its asymptotic properties. The limiting behaviour of the solution depends on how the axial period of the solution $2\pi/k$ is scaled by $Ta$. The range of $k$ from $O(1)$ to the value where it is so large that the nonlinear solution ceases to exist is covered in the analysis. The Taylor vortex solution with a wavenumber $k$ of $O(1)$ reproduces surprisingly well the properties of the large-scale coherent structures in the classical turbulence regime. \textcolor{black}{The numerical results also suggest that there may be some connection between the onset of the ultimate turbulence regime and the high wavenumber solutions, although the precise role of the solutions in the dynamics needs to be clarified in the future.}

\textcolor{black}{Our results provide evidence that the strategy of taking a simple solution from the dynamics and applying the matched asymptotic expansion for it can be used even for fully developed high-Reynolds-number turbulence to some extent.} This finding offers great hope for the first-principles explanation for turbulent momentum and heat transport that is still poorly understood.

\subsection{Summary of the asymptotic properties of the solutions}
When $k=O(1)$, the asymptotic solution consists of an inviscid core and a viscous boundary layer enveloping it. In the core region, the Prandtl-Bachelor theorem imposes strong restrictions on the structure of the flow field. The asymptotic scaling of the wind Reynolds number $Re_w=O(Ta^{1/2})$ is consistent with the turbulent observations \textcolor{black}{(Huisman et al. 2012, Ostilla et al. 2013).} On the other hand, the viscous boundary layer can be divided into a wall boundary layer and a plume. According to the asymptotic analysis, it is the Coriolis force acting on the plume that drives the overall flow. The theoretical Nusselt number for the $k=O(1)$ solutions is $Nu=O(Ta^{1/4})$, and the DNS data suggests that the same $Nu$ scaling can be applied for the classical turbulence.

The Nusselt number of the solution reaches its maximum when $k=O(Ta^{2/9})$. This asymptotic state, which we call the first peak state, has a structure similar to the $k=O(1)$ flow in a neighbourhood of the walls. The near-wall boundary layer becomes thinner than the $k=O(1)$ case, so naturally, the Nusselt number is increased. At the same time, viscous effects in the plume \textcolor{black}{dominate} Coriolis forces. This means that the Coriolis force must act in the core to drive the vortex, and this condition determines the scaling $Nu=O(Ta^{1/3})$ and $Re_w=O(Ta^{4/9})$. \textcolor{black}{For large wavenumber solutions, the radial velocity is much larger than the axial velocity, so the scaling of $Re_w$ is the same for the definitions used by Huisman et al. (2012) and Ostilla et al. (2013).}

There is another local maximum for the Nusselt number, which occurs at $k=O(Ta^{1/4})$. In this second peak state, the nonlinear interaction of the axially fluctuating component disappears in the core. As a result, only a single Fourier mode and mean flow play a major role.  To connect this core flow towards the cylinder walls, three near-wall layers are necessary. The bottom boundary layer adjacent to the cylinder wall is needed in order to satisfy the no-slip boundary conditions, and the top boundary layer must appear in order to attenuate the Fourier harmonics towards the core region. These two boundary layers of different natures can be matched through the middle boundary layer. The Nusselt number and wind Reynolds number scalings are $O(Ta^{3/10})$ and $O(Ta^{2/5})$, respectively, ignoring a minor logarithmic correction. The Coriolis force is a leading-order effect in all regions except the bottom boundary layer.

The second peak state only appears when the scaled wavenumber $kTa^{-1/4}$ is smaller than the critical value $K_{\infty}$, which can be determined analytically. As the wavenumber $k$ is further increased, the Taylor vortex undergoes a transitional state similar to that analysed by Denier (1992) and then becomes a laminar flow. The Nusselt number of the transitional state is $O(Ta^0)$ and can be found analytically, which agrees well with the numerical results.

In summary, as $k$ is reduced from the linear critical point, the Taylor vortex solution develops according to the following scenario. First, vortices grow from near the inner cylinder in the transitional state. When the vortices reach the outer cylinder, the flow field becomes the second peak state. A further decrease in $k$ increases the thickness of the outer boundary layer, where Fourier harmonics are seen. The first peak state appears when the outer boundary layers on both sides come into contact. 
\textcolor{black}{The axial scale of this state is sufficiently large for the separation of the inviscid region and the plume to be noticeable in the near-wall regions. When $k=O(1)$, the inviscid region extends across the gap, and the scaling for the largest-scale vortices is established.}

\textcolor{black}{How the above steady solutions relate to the turbulence dynamics is a natural question; Kooloth et al. (2021) investigated this problem for two-dimensional RBC and their results may also be applicable to our case. Their key observation is that local structures in the turbulent dynamics can be approximated by the relatively short-wavelength solutions if the wavelength is suitably chosen. 
The detailed comparison of the dynamics and the solutions revealed that local structures with a wide range of wavelengths are embedded in turbulence. 
It is still unknown how often the solution of each wavelength appear in the dynamics. But if all solutions are equally likely to appear, then the solutions with the largest $Nu$ scaling will have a greatest impact on the average value of $Nu$ in the dynamics. 
}

An interesting aspect of the above asymptotic theories is that it shows what the mean flow $\overline{v}$ in the core will look like. When $k$ is not too large, the mean angular momentum $r\overline{v}$ becomes a constant in the core region. In the $k=O(1)$ case, this is a consequence of the Prandtl-Batchelor theorem. The mean angular momentum profile becomes nonuniform when $k=O(Ta^{1/4})$, but instead, an analytical solution can be derived up to an unknown constant. The time-averaged turbulence data support the former uniform angular momentum law (Wattendorf 1935; Taylor 1935; Smith \& Townsend 1982; Lewis \& Swinney 1999; Ostilla-M\'onico et al. 2014b). The fact that $\overline{v}$ is proportional to $1/r$ implies that the mean flow is irrotational. By taking the narrow gap limit, it can be seen that the argument here explains why zero absolute vorticity states are commonly found in rotating parallel shear flows (see Johnston et al. 1972, Tanaka et al. 2000, Suryadi et al. 2014, Kawata et al. 2016).

We further remark that the uniform angular momentum states are a natural generalisation of the uniform momentum states often observed in non-rotating shear flows. In particular, the staircase-like uniform momentum zones ubiquitously seen in the near-wall turbulent boundary layer have long attracted much attention (Meinhart \& Adrian 1995; de Silva, Hutchins \& Marusic 2016). To explain this phenomenon, Montemuro et al. (2020) and Blackburn et al. (2021) attempted to construct a theory for homogeneous shear flows. At the heart of these theories is the Prandtl-Batchelor theorem, which was first proposed to apply for the uniform momentum state by Deguchi \& Hall (2014a) for a steady solution of plane Couette flow. It is however not yet clear how such a core structure interacts with the near wall flow and deduces the scaling of the momentum transport in Reynolds number.

\subsection{Link to the RBC studies}
The asymptotic analyses of the first and second peaks presented in this paper can also be applied to \textcolor{black}{the roll-cell solutions of RBC (regardless of whether the boundary is slip or no-slip)} and therefore provide a theoretical explanation for the scaling obtained by the numerical computation in Waleffe et al. (2015) and Sondak et al. (2015). Note, however that for RBC, the transitional state only exists in the vicinity of the bifurcation point in the parameter space (Blennerhassett \& Bassom 1994). This is due to the fact that in RBC, the instability occurs uniformly in the flow, whereas in TCF, the centrifugal instability is usually most pronounced near the inner cylinder.

\textcolor{black}{The computation of the roll-cell solution branch by Wen et al. (2022) reaches $Ra=10^{14}$, which is close to the experimentally feasible limit. In their numerical result the exponent for $k$ at the first peak is closer to 1/5 than 2/9. However, their $Nu$ exponent is also slightly off from 1/3 so a calculation with a larger $Ra$ would be necessary to obtain the exponents accurately. Of course, the possibility that an unknown asymptotic state exists cannot be ruled out, but it appears from the author's experience that it is difficult to construct new sensible asymptotic theories.
}

It is worth mentioning that Iyer et al. (2020) obtained the Nusselt number exponents 0.29 and 0.331 for moderate and high Rayleigh number regimes, respectively, for RBC in a vertically elongated tank. These exponents are close to those of the second and first peaks. The scaling of the first peak matches with the so-called classical scaling, but our derivation significantly differs from that of Priestley (1954), Malkus (1954), Grossman \& Lohse (2000). Recently, Kawano et al. (2021) %proposed the derivation of the scaling laws 
%\textcolor{black}{from physical insights into the governing equations}. 
\textcolor{black}{derived the scaling laws by considering the naive balance of each term in the governing equations within the boundary layer.}
Although they do not assume the two-dimensionality of the flow, the scaling is consistent with our first peak asymptotic state. In particular, the wind Reynolds number scaling $Re_w=O(Ta^{4/9})$ we found is in agreement with the results by Grossman \& Lohse (2000) and Kawano et al. (2021). As pointed out by the latter paper, the exponent \textcolor{black}{$4/9\approx 0.444$} is close to 0.458 observed in the high Rayleigh number DNS by Iyer et al. (2020). \textcolor{black}{Their wind Reynolds number is defined by the root-mean square of the three velocity components normalised by viscous velocity scale $\nu/H$, where $H$ is the vertical length of the tank, and thus comparable with our results. Iyer et al. (2020) used a computational domain where the horizontal scale is 1/10 of the vertical scale, and it therefore makes sense that the result is related to our large $k$ solutions.}

\textcolor{black}{
The exponent of the Nusselt number obtained in Iyer et al. (2020) is smaller than the 0.38 obtained by He et al. (2012). The tank aspect ratio might be one of the causes of this difference. However, there are only a few reports of $Nu$ exponents exceeding 1/3 in RBC, and the true nature of ultimate RBC scaling is still a matter of active debate. Wen et al. (2022) summarised the $Nu$ data of turbulent RBC and found that they were all smaller than those of the optimumised steady roll cell solution corresponding to the first peak state.
Zhu et al. (2018) restricted the DNS of RBC to two dimensions and increased the Rayleigh number to $10^{14}$ hoping to see the ultimate turbulence scaling. However, their numerical data are scaled by the exponent $\beta=1/3$, as pointed out by Doering et al. (2019). 
Even without restricting to two-dimensional steady states, the existence of simple solutions with an exponent greater than $1/3$ is still unknown. Motoki et al. (2018) computed a three-dimensional steady solution of RBC, but the $Nu$ scaling was similar to that seen in Waleffe et al. (2015) and Sondak et al. (2015). 
}

%Past literature on natural convection between no-slip walls supports that a similar restriction in the flow changes the nature of turbulence. 

\subsection{What is missing in the theoretical results?}
%---- ultimate -----
%Our analysis shows that the exponent $\beta$ of the Nusselt number scaling $Nu \propto Ta^{\beta}$  is at most 1/3 for the Taylor vortex. 
\textcolor{black}{Unlike RBC, the emergence of the Nusselt number exponent $\beta=0.38$ is well-established in the ultimate turbulence regime of TCF.
%exponent is smaller than that obtained for the ultimate turbulence, $\beta=0.38$.
Therefore, the predictions by the first peak state} should eventually deviate \textcolor{black}{from the turbulent measurement} as the Taylor number increases. In the ultimate turbulence, a mix of large-scale and small-scale structures was observed, so in a sense, it is natural that this difference should appear. Experiments have shown that large-scale structures with scaling of $Re_w=Ta^{1/2}$ can be observed even in the ultimate turbulence regime (see \textcolor{black}{Huisman et al. 2012}). This suggests that the $k=O(1)$ state studied in section 4 still plays some role in the dynamics, but as we have seen in our analysis, it cannot \textcolor{black}{efficiently transport angular momentum.} The reason for the inefficiency is the Prandtl-Batchelor homogenisation of $\Gamma$ as remarked just below (4.15). Thus it would be a natural idea to introduce small-scale vortices in the core to enhance transport. Note however that vortices of about Kolmogorov microscale $Re_w^{-1/2}$ are unlikely to show any improvement. The typical velocity scale $Re_w^{1/2}$ (see Deguchi 2015, for example) yields $O(\widetilde{\Gamma}|_c)=O(Ta^{-1/2}Re_w^{1/2})$, so from the transport balance $Nu=O(Re_w \widetilde{\Gamma}|_c)$ derived by (4.15) the scaling remains $Nu=O(Ta^{1/4})$. On the other hand, the azimuthal velocity perturbations generated by the first and second peak states are larger and may improve transport. If instead $O(\widetilde{\Gamma}|_c)=O(Ta^{-1/10})$ obtained in the second peak analysis is used, the resultant scaling $Nu=O(Ta^{0.4})$ is close to the experimental observation, though of course it is not known whether such a flow structure is possible in the asymptotic expansion framework.

\textcolor{black}{Another possible reason for the deviation of the $Nu$ scaling would be the axisymmetric restriction imposed for the Taylor vortex. 
When the three-dimensionality of TCF is allowed, feedback through Reynolds stresses appears from the non-axisymmetric component to the axisymmetric component. This feedback effect is a key process in self-sustaining the coherent structures in non-rotating shear flows (Waleffe 1997), and recently Dessup et al. (2018), Sacco et al. (2019) attempted to extend this idea to TCF. Our results suggest that the feedback is unimportant for classical turbulence, \textcolor{black}{though} it may not be so for ultimate turbulence. }

\textcolor{black}{
Minor modifications to the theories obtained in this paper using the existing asymptotic results would not fully explain the scaling of the ultimate turbulence.
For example, azimuthal and time-dependent waves can be incorporated into our axisymmetric asymptotic theories, using the method used in Hall \& Sherwin (2010), but this does not change the $Nu$ scaling.
The crux of this problem would be that all existing nonlinear asymptotic theories for shear flows do not much change the scaling of the wall shear rate from that for laminar flows. }
%Shear flow ... friction factor ...

\textcolor{black}{
Experiments and numerical simulations have shown that there is universality in the friction factor of various high Reynolds number shear flows (see Orlandi et al. (2015) for example). 
%However, a simple solution to reproduce this universality has not been found so far.
No rational asymptotic theory has been found to explain this friction factor.
If a new simple solution that reproduces the universal friction factor scaling can be found, it might serve an important hint for understanding the ultimate turbulence of TCF as well. 
%\textcolor{black}{three-dimensional ... analogy not possible ...}
For example, if one assumes the empirical Blasius friction law (friction factor $\propto Re^{-1/4}$), it implies the emergence of near wall boundary layer of thickness $O(Re^{-3/4})$. In terms of the Taylor number the boundary layer thickness is $O(Ta^{-3/8})$, so the Nusselt number exponent 
is $3/8=0.375$, which is not too far from 0.38.
The similarities between the friction factor of TCF and other shear flows are also noted by Lathrop et al. (1992).
%$Nu=O(Ta^{3/8})=O(Ta^{0.375})$.
%Whether there is a minimum azimuthal scale at which the ultimate turbulence can be sustained is an interesting question. This could be ascertained by varying the size of the periodic box in the DNS. Finding a relatively simple solution that is representative of the minimal ultimate turbulence would be the first step towards a further understanding of the flow scaling. 
If the hypothesis that the ultimate turbulence of TCF shares a common mechanism with that of non-rotating shear flows is true, then %it would be fair to say that the TCF-RBC analogy is broken in the extreme parameters.
this could settle the debate on whether the TCF-RBC analogy is valid in the extreme parameters.
}

\textcolor{black}{
Finally, we remark that the asymptotic structure discussed in this paper is valid when $R_i$ is smaller than or equal to $O(|R_o|)$, and hence it does not cover the whole parameter space. When the cylinders are counter-rotating, on the neutral curve the scaling $R_i=O(|R_o|^{5/3})$ is established, as noted by Donnelly \& Fultz (1960) using a dimensional argument, by Esser \& Grossman (1996) using a heuristic extension of Rayleigh's stability condition, and by Deguchi (2016) using an asymptotic analysis. It is well-known that non-axisymmetric disturbances are important near the neutral curve and nonlinear patterns such as spirals and ribbons emerge. 
Furthermore, some solution branches can be continued in the subcritical parameter regime, and in this case non-axisymmetry of the flow is critically important to sustain the nonlinear flow; see Deguchi et al. (2014), Wang et al. (2022). Developing a nonlinear asymptotic theory capable of describing the above non-axisymmetric phenomena is another interesting piece of future work.
}

%Non-axisymmetric solutions.
%Spiral and ribbons would be possible.

%We have not studied very counter-rotating cases ... but ...Donnelly \& Fultz (1960), Esser \& Grossman (1996), Deguchi (2016).

\textcolor{black}{The author thanks Dr Ostilla-M\'onico for sharing his DNS data, and Dr David Goluskin for inspiring discussion. This work was supported by Australian Research Council Discovery Project DP230102188.}

The author reports no conflict of interest.

\appendix

\section{The narrow-gap limits}

First, we shall derive the narrow gap limit of the G\"ortler type. Here we focus on the stationary outer cylinder case studied by Denier (1992); see Deguchi (2016) for general cases. Before taking the limit, we rewrite the system (\ref{axieqs}) using $V=v/R_i$  and $y=r-r_i$. The boundary conditions become
\begin{eqnarray}
(u,V,w)=(0,0,0)\qquad \text{at} \qquad y=1,\\
(u,V,w)=(0,1,0)\qquad \text{at} \qquad y=0.
\end{eqnarray}
While taking the limit $\eta \rightarrow 1$, or equivalently $r_i\rightarrow \infty$, we assume that the transformed velocity components and its derivatives are all $O(r_i^0)$. Keeping $T=2R_i^2/r_i$ as $O(r_i^0)$ to balance the Coriolis term, the limiting flow is governed by
\begin{eqnarray}
DV = \triangle V,\qquad D\omega=\triangle \omega +TV \partial_z V,
\end{eqnarray}
and $\partial_y u+\partial_z w=0$. Here $\omega=\partial_z u-\partial_y w, D=\partial_t+u\partial_y+w\partial_z, \triangle=\partial_y^2+\partial_z^2$. 
%The narrow gap limit implies $r_i\rightarrow \infty$ and $R_i=O(r_i^{1/2})$ must be satisfied.

For the rotating plane Couette flow (RPCF) limit, on the other hand, the radial coordinate is shifted as $y=r-r_m$ so that the origin is at the mid gap. Furthermore, we take the new azimuthal velocity $V$ to satisfy  $(v-v_b)=G(V+y)$, which means that a constant $G$ scales the disturbance. Here we expect that the base flow of the transformed system becomes almost a linear profile when the gap is narrow.
The boundary conditions become
\begin{eqnarray}
(u,V,w)=(0,-1/2,0)\qquad \text{at} \qquad y=1/2,\\
(u,V,w)=(0,1/2,0)\qquad \text{at} \qquad y=-1/2.
\end{eqnarray}
The constant $G$ is chosen as $G=-(rv_b)'/r$, and the key assumption to take the system to RPCF is $O(G)\ll O(v_b)$. Keeping $T=2Gv_b(r_m)/r_m$ as $O(r_m^0)$, it is straightforward to show that in the narrow gap limit, the system (\ref{axieqs}) reduces to
\begin{eqnarray}
DV =\triangle V,\qquad 
 D\omega=\triangle \omega+T\partial_z V.
\end{eqnarray}
and $\partial_y u+\partial_z w=0$. 
The derivation of RPCF here is mathematically simpler than the common one (see Drazin \& Reid 1981), though the physical motivation would be a bit harder to see.

\textcolor{black}{
The assumption $O(G)\ll O(v_b)$ used in the RPCF limit is equivalent to $O(R_o-R_i)\ll O(R_i)$, meaning that the inner and outer cylinders rotate at approximately the same speed. Thus when considering a coordinate system rotating at the average speed of the cylinders, the Coriolis force acting on the fluid is uniform.
% and an analogy can be made by considering it as a buoyancy force.
%Thus the instability occurs uniformly across the gap
The uniform force produces an equivalent effect to buoyancy, allowing us to use a perfect correspondence with RBC. Of course, in the general case this force is not uniform, which is why in the Denier's case the instability grew from near the inner cylinder (see figure 14 also). 
%}
%\textcolor{black}{
}
\textcolor{black}{
Note that the forcing term in the general case corresponds to the third term on the right-hand side of (4.1b), and strictly speaking this is not a Coriolis force anymore. 
In fact, the definition of the Coriolis force changes depending on which rotational coordinate system is chosen, and hence it cannot be well defined when the inner and outer cylinders differentially rotate. The forcing terms are produced by the change in the direction of the base flow (i.e. the linear terms in the $r^{-1}v^2$ and $r^{-1}uv$ terms in (2.1)). 
They are not centrifugal force either - centrifugal force can be included into the pressure gradient term and does not affect the dynamics.
}

%the flow parameters are close to the Rayleigh line $R_o=R_i$. 
%Note that this Coriolis force is defined in a frame that rotates with the parallel plates.

%Near the Rayleigh line, instability occurs uniformly everywhere in the gap, which is why full correspondence with the RB is possible.

%Here $r_m$ is large, $v_b(r_m)\sim R_i \sim R_o$. 
%So the assumption implies $G\sim r_m/R_i \ll R_i$. 

%In TCF the RPCF limit is special
%Rayleigh condition uniformly violated ...
%Only this case we have exact analogy to RB.

The difference between the two limits also affects the azimuthal development of the flow, if any. For the G\"ortlier type limit, the scaling factor of the azimuthal velocity is large, and therefore the flow must have a larger length scale than the gap in the azimuthal direction. \textcolor{black}{In contrast,} for the RPCF type limit, we can make the scaling factor $G$ to be $O(1)$ when $R_i=O(r_m)$. In this case, the azimuthal length scale of the flow is comparable to the gap.

\section{The transitional states with outer cylinder rotation}

This section discusses what modifications are required if $\Gamma_o\neq 0$ in section 6.1.
The effect of the outer cylinder rotation appears in the mean flow layer solution so that (\ref{transmeanB}) must be replaced by
\begin{eqnarray}
\overline{\Gamma}=B(r_o^2-r^2)+\Gamma_o+\cdots,\qquad N_0=2(r_o^2 B+\Gamma_o).\label{transmeanB2}
\end{eqnarray}
Noting that in the core we can still use (\ref{transmeanA}) and (\ref{transA}),
%For $r>r_c$, the flow is unforced. Thus $\overline{\Gamma}=C(r_o^2-r^2)+\Gamma_o$.
the continuity of $\overline{\Gamma}$ and $r^3(r^{-2}\overline{\Gamma})_r$ at $r=r_c$ is satisfied if
\begin{eqnarray}
\sqrt{\frac{A-r_c^4}{4T_0}}=B(r_o^2-r_c^2)+\Gamma_o,\label{Apeq1}\\
\frac{A}{\sqrt{T_0(A-r_c^4)}}=2 (Br_o^2+\Gamma_o).\label{Apeq2}
\end{eqnarray}
Eliminating $B$ from those equations we get
\begin{eqnarray}
\frac{A}{r_o^2}-r_c^2=\frac{2\sqrt{T_0}}{r_o^2}\sqrt{A-r_c^4}\Gamma_o\label{rcT0}
\end{eqnarray}
which links $r_c^2$ and $T_0$.

%$r_c(T_0)$ is continuous. $r_c(T_1)=r_i$, $r_c(T_{\infty})=r_o$
Now let us suppose $r_c=r_i$ when $T_0=T_1$ in the above equation. Then the scaled wavenumber at the linear critical point can be easily found as
\begin{eqnarray}
K_1=T_1^{-1/4}, \qquad T_1=\frac{r_i^2(r_o^2-r_i^2)}{4(\Gamma_i^2-\Gamma_i\Gamma_o)}.\label{ApK1}
\end{eqnarray}
Likewise setting $r_c=r_o$ and $T_0=T_{\infty}$ in (\ref{rcT0}),
\begin{eqnarray}
\sqrt{4T_{\infty}\Gamma_i^2+r_i^4-r_o^4}\left (\sqrt{4T_{\infty}\Gamma_i^2+r_i^4-r_o^4}-\sqrt{4T_{\infty}}\Gamma_o \right )=0.\label{T0T0T0}
\end{eqnarray}
From this equation the critical scaled wavenumber where the $Nu$ blows up can be found as
\begin{eqnarray}
K_{\infty}=T_{\infty}^{-1/4}, \qquad T_{\infty}=\left \{
\begin{array}{c}
\frac{r_o^4-r_i^4}{4(\Gamma_i^2-\Gamma_o^2)}~~~\text{if}~~~\Gamma_i>\Gamma_o> 0,\\
\frac{r_o^4-r_i^4}{4\Gamma_i^2}~~~\text{if}~~~\Gamma_o\leq 0.
\end{array}
\right .\label{ApKinf}
\end{eqnarray}
Note that equation (\ref{T0T0T0}) may have two roots, but by assuming that $K_{\infty}(a)=T_{\infty}^{-1/4}$ is a continuous function and that $K_{\infty}\rightarrow 0$ as the Rayleigh line ($\Gamma_i=\Gamma_o$) is approached we can uniquely determine the solution (\ref{ApKinf}).

%We must set $A=r_o^4$ if $\sqrt{A-r_o^4}>2\sqrt{T_{\infty}}\Gamma_o$.
%$r_i^4-r_o^4+4T_0\Gamma_i^2>4T_0\Gamma_o^2$
%Then $T_2=\frac{r_o^4-r_i^4}{4\Gamma_i^2}$ if $\Gamma_o\leq 0$, $T_2=\frac{r_o^4-r_i^4}{4(\Gamma_i^2-\Gamma_o^2)}$ if $\Gamma_i>\Gamma_o> 0$.
%Note that at Rayleigh line, $\Gamma_i=\Gamma_o$.

To find $N_0$, 
we use the fact that the quadratic equation 
%we first use (\ref{rcT0}) to eliminate $r_c^2$ from the right hand side of (\ref{Apeq1}). Then further using (\ref{Apeq2}) to the factor $\sqrt{A-r_c^4}$ 
\begin{eqnarray}
0=\left(\frac{r_o^4}{A}-1\right )N_0^2+4\Gamma_oN_0-\left (4\Gamma_o^2+\frac{r_o^4}{T_0}\right )
%
%0=\left(\frac{r_o^4}{A}-1\right )(\frac{N_0}{2})^2+2\Gamma_o\frac{N_0}{2}-\left (\Gamma_o^2+\frac{r_o^4}{4T_0}\right ).
\end{eqnarray}
can be obtained by combining (\ref{Apeq1}), (\ref{Apeq2}), (\ref{rcT0}).
%Recall that $T_0\leq T_{\infty}$ implies $r_o^4-A\geq 0$
The solution
\begin{eqnarray}
N_0=2\frac{\sqrt{\Gamma_o^2+(A^{-1}r_o^4-1)(\Gamma_o^2+\frac{r_o^4}{4T_0})} -\Gamma_o}{A^{-1}r_o^4-1}
\end{eqnarray}
and (\ref{transN0}) yields
\begin{eqnarray}
%Nu=\frac{(1+\eta)^4}{8\eta r_i^2}\frac{\sqrt{\Gamma_o^2+(A^{-1}r_o^4-1)(\Gamma_o^2+\frac{r_o^4}{4T_0})} -\Gamma_o}{A^{-1}r_o^4-1}.\\
%
Nu(T_0)=\frac{(1+\eta)^4}{16\eta^3\sqrt{T_0}}
\frac{r_i^4+4T_0\Gamma_i^2}{r_o^4-r_i^4-4T_0\Gamma_i^2}
\left (\sqrt{\frac{r_o^4+4T_0\Gamma_o^2}{r_i^4+4T_0\Gamma_i^2}-1
} -\Gamma_o \right ),
%\frac{\sqrt{\frac{r_o^4+4T_0\Gamma_o^2}{r_i^4+4T_0\Gamma_i^2}-1} -\Gamma_o}{\frac{r_o^4-r_i^4-4T_0\Gamma_i^2}{r_i^4+4T_0\Gamma_i^2}}.
\end{eqnarray}
which is the generalised version of (\ref{transNuNu}).

%If $\Gamma_o=0$ ...
%\begin{eqnarray}
%Nu=\frac{(1+\eta)^4}{16\eta^3\sqrt{T_0}}\sqrt{\frac{r_i^4+4T_0\Gamma_i^2}{r_o^4-r_i^4-4T_0\Gamma_i^2}}.
%\end{eqnarray}
%===================================

\end{document}